\documentclass[11pt,a4paper]{article}
\usepackage[pdftex]{graphics}
\usepackage{jheppub}
\usepackage{float}
\usepackage{amsmath,amssymb,amsfonts}
\usepackage{feynmp-auto}

\usepackage{xcolor}
\usepackage{appendix}
\usepackage{graphicx}
\usepackage{bm}
\usepackage{tikz}
\usepackage[utf8]{inputenc}
\usepackage{array}
\usepackage[compat=1.1.0]{tikz-feynman}
\newcommand{\sk}[1]{ |#1] } 			
\newcommand{\ak}[1]{ |#1\rangle }		
\renewcommand{\sb}[1]{ [#1| }			
\newcommand{\ab}[1]{ \langle#1| }		

\newcommand{\SB}[1]{ [#1] }					
\newcommand{\AB}[1]{ \langle #1 \rangle }	
\newcommand{\ASB}[1]{ \langle #1 ] }		
\newcommand{\SAB}[1]{ [ #1 \rangle }		

\newcommand{\eq}{\begin{equation}}
\newcommand{\eqe}{\end{equation}}
\newcommand{\eqa}{\begin{eqnarray}}
\newcommand{\eqae}{\end{eqnarray}}

\newbox\charbox
\newbox\slabox
\def\s#1{{      
        \setbox\charbox=\hbox{$#1$}
        \setbox\slabox=\hbox{$/$}
        \dimen\charbox=\ht\slabox
        \advance\dimen\charbox by -\dp\slabox
        \advance\dimen\charbox by -\ht\charbox
        \advance\dimen\charbox by \dp\charbox
        \divide\dimen\charbox by 2
        \raise-\dimen\charbox\hbox to \wd\charbox{\hss/\hss}
        \llap{$#1$}
}}

\preprint{}

\title{On-shell approach to (spinning) gravitational absorption processes}
\author[1]{Yu-Jui Chen}
\author[1]{Tien Hsieh}
\author[1, 2]{Yu-Tin Huang}
\author[3]{Jung-Wook Kim}
\affiliation[1]{Department of Physics and Center for Theoretical Physics, National Taiwan University, Taipei 10617, Taiwan}
\affiliation[2]{Physics Division, National Center for Theoretical Sciences, Taipei 10617, Taiwan}
\affiliation[3]{Max Planck Institute for Gravitational Physics (Albert Einstein Institute), Am M\"uhlenberg 1, Potsdam 14476, Germany}  
\emailAdd{r10222095@ntu.edu.tw}
\emailAdd{allen607.0520@gmail.com}
\emailAdd{yutinyt@gmail.com}
\emailAdd{jung-wook.kim@aei.mpg.de}
\abstract{We utilize three point amplitudes with (spinning) particles of unequal mass and a graviton to capture the dynamics of absorption processes. We demonstrate that the construction can represent the spheroidal harmonics appearing in the Teukolsky equations. The absolute square of the ``Wilson coefficients'' in this effective description can be fixed by matching to the known absorptive cross-sections. As an application, 
we compute corrections to the gravitational Compton amplitude from the exchange of states corresponding to such absorption effects. In the super-extremal limit, the corrections generate the non-analytic $|a|$-dependent contribution of the Compton amplitude found in ref.\cite{Bautista:2022wjf}.  }

\begin{document}

\maketitle

\section{Introduction}
The large separation of scales, between the diameter of the inspiralling orbit, the wave-length of gravitational waves and the Schwarzschild radius, suggests that the inspiral dynamics of binary black holes are well described by effective field theories of point-particles~\cite{Goldberger:2004jt} (see
refs.\cite{Goldberger:2007hy, Foffa:2013qca, Rothstein:2014sra, Porto:2016pyg, Levi:2018nxp} for a comprehensive review). It is thus natural to consider extracting inspiral binary dynamics from the scattering amplitude of particles. Indeed such a possibility was successfully demonstrated more than half a century ago~\cite{Iwasaki:1971vb}, where the two-body conservative potential was captured from an S-matrix. In essence the scattering amplitude is governed by the same potential as the bound system, simply living above the threshold.     

With the advent of direct observation of gravitational waves from LIGO/Virgo Collaboration~\cite{LIGOScientific:2016dsl, LIGOScientific:2016aoc}, there has been a resurgence of interest exploring new approaches to the computation of classical observables from scattering amplitudes. One of the main motivation is to utilize the remarkable simplicity and hidden structures of on-shell amplitudes~\cite{Neill:2013wsa}, including the double copy relation between gauge and gravity amplitudes~\cite{KAWAI19861, Bern:2008qj}, generalized unitarity methods~\cite{Bern:1994zx, Bern:1994cg} and covariant on-shell variables~\cite{Arkani-Hamed:2017jhn}. An important framework was introduced by Cheung, Solon, and Rothstein~\cite{Cheung:2018wkq}, where one gives a matching procedure for the two-body effective field theory, whose interacting vertex is the potential, with the scattering amplitude. This led to the milestone computation of Bern, Cheung, Roiban,
Shen, Solon, and Zeng, giving the 3rd and 4th Post Minkowskian (PM) correction to the conservative potential~\cite{Bern:2019nnu, Bern:2021yeh}. See ref.\cite{Buonanno:2022pgc} and references therein for a more up to date review of extraction of conservative potential from scattering amplitudes.

The inclusion of spin-effects to the classical limit of amplitudes introduces further conceptual challenges. On the one hand there are general no-go theorems of elementary particles beyond spin-2 obstructing formulation of general spinning particles as quanta of elementary fields, and on the other hand the (quantum mechanical) spin scales with the Planck constant $\hbar$ and the na\"ive classical limit $\hbar \to 0$ removes spin degrees of freedom from the description. It was soon realised that spin degrees of freedom should be understood as the action of Lorentz generators projected onto the little group space of individual particles, and while their matrix elements scale with $\hbar$, the (representation-independent) projected Lorentz generators themselves should be considered classical~\cite{Chung:2019duq,Bern:2020buy}. This is why it is possible to extract classical spin dynamics from amplitudes of spinning particles~\cite{Guevara:2018wpp, Chung:2019duq,FebresCordero:2022jts}, albeit limited by representation-dependent identities such as the quadratic Casimir relation $\mathbb{S}^2 = s(s+1) \hbar^2$.

Therefore it is important to resolve the representation-dependent identities when determining the full spinning dynamics. A popular approach is to take the classical spin limit, where ``infinite spin'' representation limit $s \to \infty$ is taken together with the classical limit $\hbar \to 0$ so that $s\hbar$ is held fixed. There are several implementations of this approach in the literature. First of all, one can start with arbitrary spin effective field theory~\cite{Bern:2020buy, Bern:2022kto, Ochirov:2022nqz}, and fix the the relevant Wilson coefficients via matching procedures. Alternatively, one can ``bootstrap" the on-shell spinning amplitude~\cite{Chung:2018kqs,Chung:2019duq,Chen:2021kxt, Chiodaroli:2021eug, Cangemi:2022bew} and analytically continue to the infinite-spin limit (classical spin limit), or directly bootstrap the infinite-spin amplitude~\cite{Aoude:2020onz,Aoude:2022trd,Haddad:2023ylx,Aoude:2023vdk,Bjerrum-Bohr:2023jau,Bjerrum-Bohr:2023iey}.  
The classical spin limit is not needed in the worldline-based approaches, as spins are considered as classical variables in post-Newtonian and post-Minkowskian effective field theory~\cite{Porto:2005ac,Levi:2015msa,Liu:2021zxr}, while the spinning particles in the supersymmetric worldline QFT approach~\cite{Jakobsen:2021lvp,Jakobsen:2021zvh} are formally spin-$\frac{\mathcal{N}}{2}$ particles of finite quantum spin.\footnote{The WQFT approach avoids the ambiguities from finite spin representation identities because the Grassmann variables that act as ``square-root'' Lorentz generators are not constrained by the identities.
}
Given all these various approaches to spinning dynamics, let us pause and mention the elephant in the room; ``\emph{do they actually describe classical spinning black holes?}''.

While at three-points, with the spinning particle emitting a graviton, it has been shown that minimally coupled\footnote{Minimal coupling in the sense of ref.\cite{Arkani-Hamed:2017jhn}.} higher-spin particles reproduces the 1 PM observables of Kerr(-Newman) black holes~\cite{Guevara:2018wpp,Chung:2018kqs,Chung:2019yfs,Arkani-Hamed:2019ymq,Aoude:2020onz}, 
the situation is less clear for the two graviton emission, i.e. the gravitational ``Compton amplitude". First of all, the amplitudes are subject to polynomial ambiguities for $s>2$; the pole contributions to the four-point amplitude, which are dictated by the minimally coupled three-point amplitudes, mixes with polynomial terms beyond quartic order in spin (or $s>2$). This ambiguity is realised on the EFT side as Wilson coefficients in the off-shell Lagrangian that contributes to the 1 PM Compton, which starts to appear from quartic order in spin~\cite{Bern:2020buy,Bern:2022kto}. These ambiguities can be removed to some extent by requiring improved high energy behaviour and the presence of spin-shift symmetry~\cite{Aoude:2022trd,Bern:2022kto}. The latter emerged from the pattern of tensor structures appearing in the 2 PM spinning conservative potential up to quartic order in spin~\cite{Kosmopoulos:2021zoq,Aoude:2020ygw,Chen:2021kxt}, and to all order for the 2 PM Gravitational Faraday effect~\cite{Chen:2022clh}.

By nature, polynomial terms in an effective field theory should be fixed through a matching procedure, thus it is preferable to have a direct GR computation of the classical Compton amplitude, e.g. gravitational wave scattering by a spinning black hole studied using black hole perturbation theory (BHPT). This was done beautifully by Bautista, Guevara, Kavanagh and Vines in refs.\cite{Bautista:2021wfy,Bautista:2022wjf}, where it was shown that the minimal coupling Compton amplitude correctly describe spinning black holes up to quartic order in spin. 
However, at quintic order one observes terms of the form $|a| = \sqrt{- a_\mu a^\mu}$, where $a^\mu = S^\mu / m_1$ is the spin-length vector. Such terms are puzzling from the perspective of effective field theory, where all interaction terms are given as polynomials of operators generated by Taylor expansions around the background; non-analytic terms of the form $\sqrt{- a_\mu a^\mu}$ are expected to be absent. The source of such terms can be traced back to an analytic continuation performed in the calculation. More precisely, BHPT starts with a physical black-hole where $|a| < Gm_1$ for Kerr. The scattering wave solution is then analytically continued to the super-extremal limit $|a| \gg Gm_1$ to match the particle interpretation. The $|a|$ terms arise from factors of the form $\sqrt{1 - (a/Gm_1)^2}$, becoming $\sim \pm i |a|$ when analytically continued. Therefore the conservative terms with $|a|$ dependence appeared as the imaginary part in the original solution, which corresponds to absorptive effects.

That the full Compton amplitude should be privy to absorptive effects should not be surprising; the imaginary part of the forward-amplitude is the total cross-section, which includes both elastic and inelastic effects.\footnote{We ignore subtleties from IR divergences.} 
This connection was recently utilized  in ref.\cite{Jones:2023ugm} to compute absorptive corrections to impulse and mass-shifts. While the absorption cross-section starts at $\mathcal{O}(G^6)$ \cite{Starobinskil:1974nkd, Page:1976df}, thus na\"ively giving a 6 PM correction, the super-extremal limit introduces another parameter $a/Gm_1 \gg 1$ that can absorb the extra powers of $G$, and lead to additional terms in the Compton amplitude that has the same $G$ scaling as the leading term. Thus to clarify this issue, we wish to directly construct the an on-shell description of the absorption process, and connect it to the Compton amplitude.

The starting point is the three-point amplitude describing a spinning black-hole absorbing a gravitational wave and transitioning into an excited state. Descriptions working directly with infinite spin representation limits are not suitable for describing absorption/emission processes, since we need to keep track of spin variables with two widely separated scales; the spin of the Kerr black hole which scales as $s \sim \mathcal{O} (\hbar^{-1})$, and the angular momentum carried by the massless quanta which scales as $\ell \sim \mathcal{O} (1)$. To this end, we instead consider the coherent spin formalism introduced by Aoude and Ochirov~\cite{Aoude:2021oqj}; one treats the classical spin $s$ of the body as the expectation value of an ``thermodynamic ensemble'', and treat the variations in spin $\ell$ as small perturbations due to operator insertions. The separated scales do not mix in the calculations, and calculating exchange of intermediate states reduces to effective spin-$\ell$ intermediate state exchanges. 

Using the unequal-mass three-point amplitudes to model the absorptive kinematics, as was done in earlier work for Schwarzschild black-holes~\cite{Kim:2020dif,Aoude:2023fdm}, we reproduce the spheroidal harmonics of BHPT within the coherent-spin state formalism. The three-point amplitude is then determined up to ``Wilson coefficients" defined on the spheroidal harmonic basis. The integral of the square of these Wilson coefficients with respect to the spectral density, can be fixed by matching to the absorptive cross sections computed in ref.\cite{Ivanov:2022qqt} for the long wavelength limit.

While only the integrated sum of the Wilson coefficients can be fixed, this already allows us to extract how the absorptive modes in the long wavelength limit modify the gravitational Compton amplitude. 
By gluing the three-point amplitudes over propagators, we construct the part of the Compton amplitude whose imaginary part reproduces the absorptive cross-section. Taking the super-extremal limit of the black hole, we find that the contribution starts at quintic order in spin variables at 1 PM. We can directly compare with the result in~\cite{Bautista:2022wjf}. Using their notations, the quintic terms has the following coefficients: 
\begin{equation} \label{eq:Compton_coeffs}
    \begin{split}
        c_4^{(0)}&=\frac{45 K}{4 \pi },\;
        c_4^{(1)}=\frac{225 K}{16 \pi },\;
        c_4^{(2)}=\frac{195 K}{64 \pi },\\
        c_3^{(0)}&=\frac{4}{15}-\eta\frac{45 K}{4 \pi },\;
        c_3^{(1)}=-\eta\frac{315 K}{16 \pi },
        c_3^{(2)}=-\eta\frac{135 K}{16 \pi },\\
        c_2^{(j)}&=0,\;\;j=0,1,2.
    \end{split}
\end{equation}
where $K$ represents an overall factor associated with the sum over all excitation modes and $\eta=\pm1$ represents the sign ambiguity associated with the analytic continuation to super-extremal limit. The results in ref.\cite{Bautista:2022wjf} are 
\begin{equation}
    \begin{split}
        c_4^{(0)}&=\frac{64}{15}\alpha\eta,\;
        c_4^{(1)}=\frac{16}{5}\alpha\eta,\;
        c_4^{(2)}=\frac{4}{15}\eta,\\
        c_3^{(0)}&=\frac{64}{15}\alpha,\;
        c_3^{(1)}=\frac{16}{3}\alpha,\;
        c_3^{(2)}=\frac{4}{15}(1{+}4\alpha)\\
        c_2^{(i)}&=0.
    \end{split}
\end{equation}
Note that the result in \eqref{eq:Compton_coeffs} derived solely from the information of absorptive processes, can be viewed as complimentary to the results of ref.\cite{Bautista:2022wjf} which only focuses on conservative effects (of the super-extremal limit).

Interestingly, our calculations also suggest an explanation to the disparity between helicity-preserving and helicity-flipping Compton amplitudes, where $|a|$ dependence is only present in the former~\cite{Bautista:2022wjf}; there are degeneracies of intermediate excited states due to separation of even and odd parity states (also known as polar and axial perturbations in BHPT), and while they constructively interfere in the helicity-preserving case the interference is destructive in the helicity-flipping case. This is the spinning analogue of the cancellation due to electric-magnetic duality argued in ref.\cite{Jones:2023ugm}.

The paper is organised as follows. Section \ref{sec:2} describes gravitational partial waves in terms of on-shell variables. Using unequal-mass three point amplitude, we construct the on-shell tensors for spinless black holes in \ref{sec:2.1}, briefly review the coherent spin formalism~\cite{Aoude:2023fdm} in \ref{sec:2.2}, and construct the on-shell tensors for the spinning case in \ref{sec:2.3}. Section \ref{sec:3} presents calculation of absorption cross section of the spinning black hole in the on-shell formalism, where the effective Wilson coefficients are determined by matching to the black hole perturbation theory calculations. In Section \ref{sec:4} we compute the corrections to the gravitational Compton amplitude of spinning black holes in the on-shell formalism. We conclude and discuss future directions in Section \ref{sec:5}. 


\section{Gravitational partial wave as an on-shell tensor}\label{sec:2}

\begin{figure}
    \centering
\includegraphics[scale=0.4]{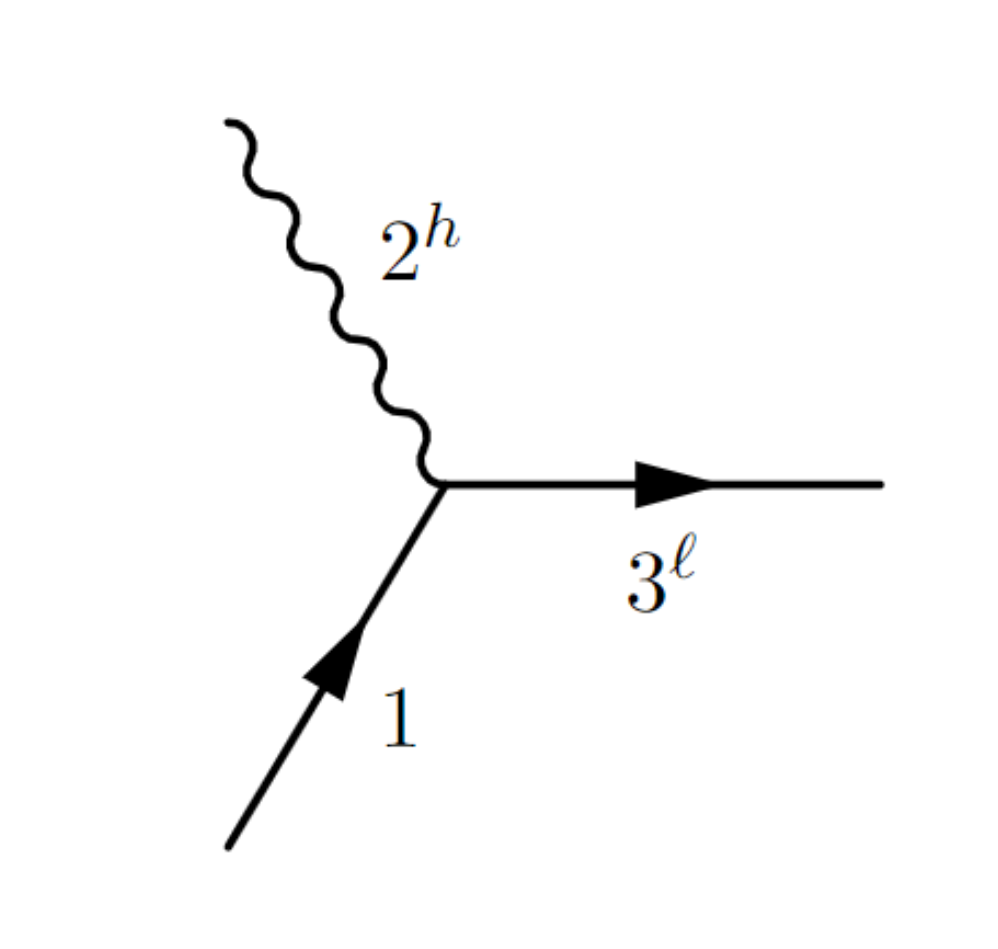}
\caption{Three point amplitude of incoming scalar massive particle $1$, helicity $h$ massless particle $2$ and outgoing spin $\ell$ massive particle $3$.}
\end{figure}
We would like to construct the spherical and spheroidal harmonics in the on-shell description. They are the angular wavefunctions of linear perturbations on Schwarzschild and Kerr black hole background respectively. 
The kinematic setup is
\begin{equation}\label{3pt_kinematic}
\begin{aligned}
    p_1^\mu &= 
    \begin{pmatrix}
	m_1 \,, &0 \,, &0 \,, &0
    \end{pmatrix},\\
    k_2^\mu &= \omega 
    \begin{pmatrix}
        {1} \,, &\sin{\theta}\cos{\phi} \,, &\sin{\theta}\sin{\phi} \,, &\cos{\theta}
    \end{pmatrix},\\
    p_3^\mu &= p_1^\mu+k_2^\mu,
\end{aligned}
\end{equation}
where $p_i$ denotes momenta of the massive particles $i=1,3$, $k_2$ is the momentum of the graviton, $m_1$ is the rest mass of particle $1$, $\omega$ is the energy of graviton, 
$\theta$ is the polar angle, and $\phi$ is the azimuthal angle. 
From the defining relations
\begin{align}
\begin{gathered}
p_{i,\alpha\dot{\alpha}}=(p_i^\mu\sigma_\mu)_{\alpha\dot{\alpha}}=\ak{{\bf i}^I}_{\alpha}\sb{{\bf i}_I}_{\dot{\alpha}}=\epsilon_{IJ}\ak{{\bf i}^I}_{\alpha}\sb{{\bf i}^J}_{\dot{\alpha}}\;\; i=1,3\;\;, \\ k_{2,\alpha\dot{\alpha}}=(k_2^\mu\sigma_\mu)_{\alpha\dot{\alpha}}=\ak{2}_{\alpha}\sb{2}_{\dot{\alpha}},
\end{gathered}
\end{align}
we parametrise the spinors of particles $1$ and $2$ as
\begin{subequations} \label{eq:spinor_param}
\begin{equation}
\begin{aligned}
    |\bf{1}^{\uparrow}\rangle_{\alpha} =|\bf{1}^{\uparrow}\rbrack^{\dot{\alpha}} =& 
    \begin{pmatrix}
	\sqrt{m_1} \\
	0 \\
    \end{pmatrix},\\
    |\bf{1}^{\downarrow}\rangle_{\alpha} =
    |{\bf1}^{\downarrow}\rbrack^{\dot{\alpha}} =& 
    \begin{pmatrix}
	0 \\
	\sqrt{m_1} \\
    \end{pmatrix},\\
|2\rangle_{\alpha} = \sqrt{2\omega}
    \begin{pmatrix}
	-e^{-i \frac{\phi}{2}} \sin \left(\frac{\theta }{2}\right) \\
	e^{i \frac{\phi}{2}}\cos \left(\frac{\theta }{2}\right) \\
    \end{pmatrix}&,\quad
    |2\rbrack^{\dot{\alpha}} =\sqrt{2\omega}
    \begin{pmatrix}
	e^{-i \frac{\phi}{2}}\cos \left(\frac{\theta }{2}\right) \\
	e^{i \frac{\phi}{2}} \sin \left(\frac{\theta }{2}\right) \\
    \end{pmatrix},\\
\end{aligned}
\end{equation}
and parametrise the spinors of particle $3$ by boosting it from its rest frame 
\begin{equation}
    \begin{split}
    |{\bf3}^{\uparrow}\rangle_{\alpha}=&
    e^{- \frac{\lambda}{2} (\hat{n} \cdot \Vec{\sigma})}
    \begin{pmatrix}
	\sqrt{m_3} \\
	0 \\
    \end{pmatrix},
    \quad
    |{\bf3}^{\downarrow}\rangle_{\alpha}=
    e^{- \frac{\lambda}{2} (\hat{n} \cdot \Vec{\sigma})}
    \begin{pmatrix}
	0 \\
	\sqrt{m_3} \\
    \end{pmatrix},
    \\
    |{\bf3}^{\uparrow}\rbrack^{\dot{\alpha}} =& 
     e^{ \frac{\lambda}{2} (\hat{n} \cdot \Vec{\sigma})}
    \begin{pmatrix}
	\sqrt{m_3} \\
	0 \\
    \end{pmatrix},
    \quad
    |{\bf3}^{\downarrow}\rbrack^{\dot{\alpha}}=
     e^{\frac{\lambda}{2} (\hat{n} \cdot \Vec{\sigma})}
    \begin{pmatrix}
	0 \\
	\sqrt{m_3} \\
    \end{pmatrix} \,,
    \end{split}
\end{equation}
\end{subequations}
where $m_3$ is its rest mass. The particle is boosted by rapidity $\lambda =\log \left(\frac{(\omega+m_1)+\omega}{m_3}\right)$ along the direction $\hat{n}= \begin{pmatrix}
        \sin{\theta}\cos{\phi} \,, &\sin{\theta}\sin{\phi} \,, &\cos{\theta}
\end{pmatrix}$ of $\vec{k}_2$.

\subsection{Spherical tensors for Schwarzschild black holes}\label{sec:2.1}
Linear perturbations on Schwarzschild black hole background can be expanded in partial waves using spin-weighted spherical harmonics ${}_{h}Y_{lm}$, which can be written as on-shell three-point amplitudes. 
For example, the three-point amplitude of massive particles $1$ and $3$ with spin $0$ and $l$, and a massless particle $2$ with helicity $h=-2$ is
\begin{equation}\label{3pt_amplitude}
\begin{aligned}
    \mathcal{A}^{\left<0\right>,-2,(J_1...J_{2l})}(\mathbf{1}^0,2^{h=-2},\mathbf{3}^l)
    =m_1\textsl{g}_{l} \mathcal{S}^{-2,(J_1...J_{2l})}(\mathbf{1}^0,2^{h=-2},\mathbf{3}^l),
\end{aligned}
\end{equation}
where the superscript $\left<0\right>$ stands for classically spinless black holes, $\textsl{g}_{l}$ is the effective ``Wilson coefficient'' of the coupling, and $\mathcal{S}^{-2,(J_1...J_{2l})}$ is the kinematic factor 
\begin{equation}\label{Kin_3pt}
   \mathcal{S}^{-2,(J_1...J_{2l})}(\mathbf{1}^0,2^{h=-2},\mathbf{3}^l)= \frac{1}{_{-2}N_{lm}}
    \langle 2'\mathbf{3'}^{(J} \rangle^{l+2}  \lbrack 2'\mathbf{3'}^{J)} \rbrack^{l-2},
\end{equation}
with symmetrised $SU(2)$ little group indices $(J_1 ... J_{2l})$ taking values $J_i=1,2= \uparrow,\downarrow$. The primed spinors are mass-rescaled dimensionless spinors, viz. 
\begin{align}
    \sb{2'}=\frac{\sb{2}}{\sqrt{m_1}} \,,\, \ab{2'}=\frac{\ab{2}}{\sqrt{m_1}} \,,\, \ak{\mathbf{3'}}=\frac{\ak{\mathbf{3}}}{\sqrt{m_3}} \,,\, \sk{\mathbf{3'}}=\frac{\sk{\mathbf{3}}}{\sqrt{m_3}} \,,
\end{align}
with $m_3 =\sqrt{2 \omega m_1+m_1^2}$ constrained by on-shell conditions in the classical limit. The denominator $_{-2}N_{lm}$ is the normalisation factor whose general form is defined as
\begin{align}
    {}_{h}N_{lm} &= (-1)^h \left( \frac{2 \omega}{m_1} \right)^l  \frac{1}{(2l)!} \sqrt{\frac{4\pi}{2l+1}} \sqrt{{(l+h)! (l-h)!}{(l+m)! (l-m)!}}
\end{align}
and $_{-2}N_{lm}$ is the special case $h = -2$. Inserting spinor parametrisations \eqref{eq:spinor_param} one can show that \eqref{Kin_3pt} reproduces spin-weighted spherical harmonics of spin-weight $-2$.
\begin{equation}\label{sphericalY}
\begin{aligned}
    \mathcal{S}^{-2,(J_1...J_{2l})}
    = {_{-2}}Y_{lm}(\theta,\phi),
\end{aligned}
\end{equation}
Generalisation to generic helicity $h$ is straightforward.
\begin{equation}
\begin{aligned}
    \mathcal{S}^{h,(J_1...J_{2l})}(\mathbf{1}^0,2^{h},\mathbf{3}^l)
    = \frac{1}{_{h}N_{lm}}\langle 2'\mathbf{3'}^{(J} \rangle^{l-h}  \lbrack 2'\mathbf{3'}^{J)} \rbrack^{l+h} = {_h}Y_{lm}(\theta,\phi) \,.
\end{aligned}
\end{equation}

\subsection{Review of the coherent spin formalism}\label{sc}\label{sec:2.2}
As mentioned in the introduction, while it is possible to describe classical spin using the spinor-helicity formalism by taking the classical spin (or infinite spin representation) limit, this approach is not very useful for describing absorption/emission processes of Kerr black holes. This is because the spinor-helicity variables organise states in irreducible representations of the little group, therefore the scale separation of black hole's classical spin $s \sim \mathcal{O} (\hbar^{-1})$ and massless particle's angular momentum $\ell \sim \mathcal{O} (1)$ is not fully utilised. A more useful approach is the coherent spin formalism introduced in ref.\cite{Aoude:2021oqj}, which we will review shortly. The usefulness can be traced back to the different treatment of angular momenta at different scales, where the classical spin is treated as the expectation value of the ``thermal ensemble'' and massless particle's angular momentum is treated as operators acting on the ensemble. We remark that conventional QFT tools can be readily extended to coherent spin states, and present a BCFW recursion~\cite{Britto:2005fq} calculation using coherent spin states as an example in appendix \ref{equalm_Com}.

The formalism starts out by embedding the rotational part of the Lorentz algebra $\mathfrak{su}(2)$ into two copies of the Heisenberg algebra $\mathfrak{h} \oplus \mathfrak{h}$, which is the algebra of isotropic 2-dimensional simple harmonic oscillator. The algebra is constructed from mode operators of the oscillator satisfying the relation 
\begin{equation}
\begin{aligned}
    \left[ \hat{a}^I, \hat{a}^{\dag}_J \right] = \delta^I_J
\end{aligned}
\end{equation}
where $I,J = 1,2$ denotes which copy of the Heisenberg algebra the mode operator comes from, which will be mapped to $SU(2)$ indices of the rotation group. The $\mathfrak{su}(2)$ generators are constructed from the mode operators using Pauli matrices as the coefficients\footnote{The defining index positioning of the Pauli matrices is $\left[\sigma^i\right]^{I}_{\;\;J}$. Indices are raised and lowered with $\epsilon^{IJ}$ and $\epsilon_{IJ}$ with $\epsilon^{12} = -\epsilon_{12} = +1$. For example, $\left[\sigma^i\right]_{KJ} = \epsilon_{KI} \left[\sigma^i\right]^{I}_{\;\;J}$ and $\left[\sigma^i\right]^{IK} = \epsilon^{KJ} \left[\sigma^i\right]^{I}_{\;\;J}$. 
}
\begin{equation}\label{spin operator}
\begin{aligned}
    S^i = \frac{\hbar}{2} \hat{a}^{\dag}_I {[\sigma^{i}]^{IJ}} \hat{a}_J.
\end{aligned}
\end{equation}
It is easy to verify that the operators satsify the $\mathfrak{su}(2)$ algebra
\begin{equation}
\begin{aligned}
    \left[ S^i, S^j \right] = i\hbar \epsilon^{ijk} S^k.
\end{aligned}
\end{equation}
The matrix element of the angular momentum operator is
\begin{equation}
\begin{aligned}
    \langle s,(I_1...I_{2s})| S_i |s,(J_1...J_{2s}) \rangle 
    = \frac{-\hbar}{2} (2s) {{[\sigma_i]}^{(I_1}}_{(J_1} \delta^{I_2}_{J_2} ... \delta^{I_{2s-1})}_{J_{2s-1})}
\end{aligned}
\end{equation}
where the spin-$s$ states are constructed from the harmonic oscillator vacuum $|0\rangle$ as
\begin{equation}
\begin{aligned}
    |s,(I_1...I_{2s}) \rangle = \frac{1}{\sqrt{(2s)!}} \hat{a}^{\dag}_{I_1}...\hat{a}^{\dag}_{I_{2s}} |0\rangle \,,\, \langle s,(I_1...I_{2s})|s',(J_1...J_{2s'}) \rangle =\delta^{s}_{s'} \delta^{(I_1}_{(J_1} ... \delta^{I_{2s})}_{J_{2s})}.
\end{aligned}
\end{equation}
The \emph{coherent spin states} are constructed from coherent states of simple harmonic oscillators
\begin{equation}\label{coherent spin state}
\begin{aligned}
    | \alpha \rangle
    =& e^{-\frac{1}{2}\tilde{\alpha}_J\alpha^J} e^{\alpha^I \hat{a}^{\dag}_I} |0 \rangle
    = e^{-\frac{1}{2} ||\alpha||^2} \sum_{2s=0} ^{\infty} \sum_{I_1,...,I_{2s}=\uparrow,\downarrow} \frac{(\alpha^{I})^{2s}}{\sqrt{(2s)!}} |s,(I_1...I_{2s}) \rangle,
\end{aligned}
\end{equation}
where $(\alpha^{I_i})^{2s} = \alpha^{I_1} \alpha^{I_2}...\alpha^{I_{2s-1}}\alpha^{I_{2s}}$, which can be thought of as a weighted sum of highest weight states ($| s, s_z = + s \rangle$) rotated to a certain direction. The overall scaling factor is needed to normalise the coherent spin states, i.e. $\langle \alpha | \alpha \rangle = 1$.
The coherent spin states have the property that the uncertainty of the spin is minimised. From the definition of the $\mathfrak{su}(2)$ generator \eqref{spin operator} we get the expectation value for the spin
\begin{equation}\label{expectation value}
\begin{aligned}
    \langle \alpha| S_i | \alpha \rangle
    = \frac{\hbar}{2} \left[ \tilde{\alpha}_I {{[\sigma_i]}^{IJ}} \alpha_J \right] \,,
\end{aligned}
\end{equation}
and we scale $\tilde{\alpha}_I \sim \alpha^I \sim \hbar^{-1/2}$ for the classical spin limit. The variance of the spin can be computed from
\begin{equation}
\begin{aligned}
    \langle \alpha| S_i S_j | \alpha \rangle
    =& \langle \alpha| S_i | \alpha \rangle  \langle \alpha| S_j | \alpha \rangle + \frac{\hbar^2}{4} \left[ \delta_{ij} (\tilde{\alpha}_I \alpha^I) +i \epsilon_{ijk} (\tilde{\alpha}_I {{[\sigma_k]}^{IJ}} \alpha_J) \right]\\
    =&\langle \alpha| S_i | \alpha \rangle  \langle \alpha| S_j | \alpha \rangle+\mathcal{O}(\hbar),
\end{aligned}
\end{equation}
and the variance vanishes in the classical limit $\hbar\rightarrow 0$. The spin vector can be covariantised into the Pauli-Lubanski pseudovector as
\begin{equation}\label{spin operator}
\begin{aligned}
    S_p^\mu = \frac{\hbar}{2} \hat{a}^{\dag}_I {[\sigma_p^{\mu}]^{IJ}} \hat{a}_J,
\end{aligned}
\end{equation}
using the momentum-dependent $\sigma$-matrices
\begin{equation}
    {[\sigma_{p}^{\mu}]^{IJ}} = \frac{1}{2m_p} \left(  \langle  p^I| \sigma^\mu |p^J\rbrack + \lbrack p^I| \bar{\sigma}^\mu |p^J\rangle \right),
\end{equation}
where $m_p$ is the mass of the particle with momentum $p^\mu$, and $\{ |p_J\rbrack \,,\, |p_J\rangle \}$ are the associated massive spinors. The $\mathfrak{su}(2)$ algebra is covariantised to
\begin{equation}
\begin{aligned}
    \left[ S_p^\mu, S_p^\nu \right] = \frac{i\hbar}{m} \epsilon^{\mu\nu\rho\sigma} p_{\rho} S_{p\;\sigma}.
\end{aligned}
\end{equation}
It is useful to define the commonly used \emph{spin-length vector}, e.g. as in BHPT. 
\begin{equation}\label{spin_vector}
\begin{aligned}
    {a}_{p}^\mu 
    = \frac{1}{m_p} \langle \alpha| S_{p}^{\mu}|\alpha \rangle \,.
\end{aligned}
\end{equation}
\subsection{Spheroidal tensors for Kerr black holes}\label{sec:2.3}
Spin-weighted spheroidal harmonics ${_h}S_{lm}(a\omega,{\theta},\phi)$ are defined as the solution to the angular Teukolsky equation
\begin{equation}\label{angular equation}
\begin{aligned}
    \left[ \frac{\partial}{\partial X} (1-X^2) \frac{\partial}{\partial X} \right] {_h}S_{lm}
    +\left[ (a\omega)^2X^2 -2a\omega h X -\frac{(m +h X)^2}{1-X^2} +h + {}_h A_{lm}(a\omega) \right]  {_h}S_{lm}
    = 0
\end{aligned}
\end{equation}
where $(\theta, \phi)$ are angular coordinates, $X=\cos{\theta}$, ${}_h A_{lm}(a\omega)$ is the eigenvalue for linearised wave equations. The label $l$ has an interpretation as the orbital angular quantum number, as the spheroidal harmonics reduce to spherical harmonics in the $a\omega \to 0$ limit; ${}_{h}S_{lm} (0,{\theta},\phi) = {}_h Y_{lm} ({\theta},\phi)$. While the closed-form formula for spheroidal harmonics ${}_{h}S_{lm} (a\omega, {\theta},\phi)$ is not known, its $a\omega$ expansion is known in terms of spherical harmonics. We attempt to construct a spinor-based expression for spheroidal harmonics utilising its $a\omega$ expansion.

Conforming with conventions, we align the black hole spin $\textsl{a}$ along the $z$-axis
\eqref{spin_vector}
\begin{equation}
    {a}^\mu 
    = \frac{1}{m_p} \langle \alpha| S_{p}^{\mu}|\alpha \rangle =(0,0,0,\frac{\textsl{a}}{m_p})=(0,0,0,{a}), \label{eq:spin_dir}
\end{equation}
in the rest frame of momentum $p^\mu = p_1^\mu$. Note that we have define the spin-length $a=\frac{\textsl{a}}{m_p}$. The coherent spin parameters satisfy the condition 
\begin{equation}
\begin{aligned}
    ||\alpha||^2 = \tilde{\alpha}_I(p) \alpha^I(p) = \textsl{a}.
\end{aligned}
\end{equation}
From an explicit matrix expansion of the parameters
\begin{equation}
\begin{aligned}
    \alpha^I(p_\textsl{a}') &=
    \begin{pmatrix}
	\alpha^1 & \alpha^2 
    \end{pmatrix} \,,\,
    \tilde{\alpha}_I(p_\textsl{a}') &=
    \begin{pmatrix}
	{(\alpha^1)}^\ast & {(\alpha^2)}^\ast 
    \end{pmatrix} \,,
\end{aligned}
\end{equation}
where the asterisk denotes complex conjugation, we obtain the solutions
\begin{equation}
\begin{aligned}
    \alpha^I(p_\textsl{a}') &= \sqrt{\textsl{a}}
    \begin{pmatrix}
	0 & e^{i \phi} 
    \end{pmatrix} \,,\,
    \tilde{\alpha}_I(p_\textsl{a}') &=\sqrt{\textsl{a}}
    \begin{pmatrix}
	0 & e^{-i \phi} 
    \end{pmatrix} \,,
\end{aligned}
\end{equation}
where $\phi \in \mathbb{R}$ is a phase. We can set $\phi = 0$ since the results do not depend on $\phi$.

From the kinematics given in \eqref{3pt_kinematic} we have the following building blocks for spinor expressions
\begin{equation}
\begin{aligned}
    \langle 2 \mathbf{3}^{I} \rangle \,,\, 
    \lbrack 2 \mathbf{3}^{I} \rbrack \,,\, 
    \langle 2 \mathbf{1}^{I} \rangle \,,\, 
    \lbrack 2 \mathbf{1}^{I} \rbrack \,,
\end{aligned}
\end{equation}
where we nondimensionalise the building blocks by the mass-rescaling $|{\mathbf{1}}^I\rangle\rightarrow\frac{1}{\sqrt{m_1}}|\mathbf{1}^I\rangle$, $|{\mathbf{3}}^I\rangle\rightarrow\frac{1}{\sqrt{m_3}}|\mathbf{3}^I\rangle$, $|2\rangle \rightarrow \frac{1}{\sqrt{m_1}}|{2}\rangle$, and analogously for the square spinors. We combine the spinor brackets with coherent spin parameters $\{ \alpha_I , \tilde{\alpha}_I \}$ to form the little-group-weight neutral building blocks for spin-weighted spheroidal harmonics in the $\gamma = a\omega$ expansion,
\begin{equation}\label{alpha2321alpha}
\begin{aligned}
    \tilde{\alpha}_K \lbrack 2{\mathbf3}^{K} \rbrack  \langle 2{\mathbf1}^{I} \rangle \alpha_I 
    &= -\frac{\textsl{a} \omega (1+X)}{\sqrt{m_1 m_3}}
    = -(1+X)\frac{\textsl{a} \omega}{m_1} +\mathcal{O}\left( \frac{\textsl{a} \omega^2}{m_1^2}\right) \,,\\
    \tilde{\alpha}_K \langle 2{\mathbf3}^{K} \rangle  \lbrack 2{\mathbf1}^{I} \rbrack \alpha_I 
    &= -\frac{\textsl{a} \omega (-1+X)}{\sqrt{m_1 m_3}}
    = (1-X) \frac{\textsl{a} \omega}{m_1} +\mathcal{O}\left( \frac{\textsl{a} \omega^2}{m_1^2}\right).
\end{aligned}
\end{equation}
For $\hbar$ counting, we scale the variables as
\begin{equation}
\begin{aligned}
    \textsl{a} &\rightarrow \frac{\textsl{a}}{\hbar} \,,\, {\omega} \rightarrow \hbar \omega
\end{aligned}
\end{equation}
such that $\frac{\textsl{a}\omega}{m_1}=a\omega$ remains $\mathcal{O}(\hbar^0)$ in the classical limit.

To construct an ansatz for spheroidal tensors, we reinterpret the three-point amplitude \eqref{3pt_amplitude} as a matrix element of an operator inserted between coherent spin states.
\begin{align}
\begin{gathered}
    \mathcal{A}^{\left<0\right>,-2,(J_1...J_{2l})}(\mathbf{1}^0,2^{h=-2},\mathbf{3}^l) \to \mathcal{A}^{\left<0\right>,-2,(J_1...J_{2l})}(\{ p_1; \alpha_1 \},2^{h=-2}, \{ p_3; \alpha_3 \}) \,,
    \\ \mathcal{A}^{\left<0\right>,-2,(J_1...J_{2l})}(\{ p_1; \alpha_1 \},2^{h=-2}, \{ p_3; \alpha_3 \}) = m_1 \langle \alpha_3 | \breve{g}_{l} [\hat{\mathcal{S}}^{-2,(J_1 \cdots J_{2l})}] | \alpha_1 \rangle \,,
\end{gathered}
\end{align}
where we view the operator $\hat{\mathcal{S}}^{-2, J_{2l}}$ as an operator proportional to the identity constructed from the spinor brackets $\{ \langle 2 \mathbf{3}^{I} \rangle , \lbrack 2 \mathbf{3}^{I} \rbrack , \langle 2 \mathbf{1}^{I} \rangle , \lbrack 2 \mathbf{1}^{I} \rbrack \}$, and promote the ``Wilson coefficient'' to an operator $\breve{g}_{l}$.\footnote{We promote the Wilson coefficient to an operator because dependence on the $m$ quantum number can only be introduced using the spin operator (Wigner-Eckart). Formally the operator $\breve{g}_{l}$ acts on $\hat{\mathcal{S}}$.} We recover the original expression \eqref{3pt_amplitude} when we set $\alpha_1 = \alpha_3 = 0$. Next, we generalise the operators $\breve{g}_{l}$ and $\hat{\mathcal{S}}^{-2, J_{2l}}$ to also depend on the mode operators $\{ \hat{a} , \hat{a}^\dagger \}$, which has an expansion as 
\begin{equation}\label{pre_tensor}
\begin{aligned}
    \mathcal{A}^{\left<a\right>,-2,(J_1 ...J_{2l})} &= m_1 \langle \alpha_3 | \breve{g}_{l} [\hat{\mathcal{S}}^{-2, (J_1 ...J_{2l})} ( \hat{a}^\dagger , \hat{a} )] | \alpha_1 \rangle,
    \\ &= m_1 \sum_{s=0}^{\infty} \langle \alpha_3 | (\hat{a}^\dagger_K)^{s} \breve{g}_{l} [\mathcal{S}^{-2, (J_1 ...J_{2l}),K_1 ...K_{s}, I_1 ...I_{s}} ] (\hat{a}_I)^{s} | \alpha_1 \rangle,
\end{aligned}
\end{equation}
where the spin of the coherent spin state is along the $z$-axis as in \eqref{eq:spin_dir} and $J$'s are symmetrised free little group indices carried by the leg 3. We use the property of coherent spin states,
\begin{align}
    \hat{a}_I | \alpha \rangle = \alpha_I | \alpha \rangle \,,\, \langle \alpha_1 | \alpha_2 \rangle = e^{-\frac{||\alpha_1||^2 + ||\alpha_2||^2}{2} + \tilde{\alpha}_{1I} \alpha_2^I} \,,
\end{align}
to write the amplitude as
\begin{align}
    \mathcal{A}^{\left<a\right>,-2,(J_1 ...J_{2l})} &=  m_1g_{lm} e^{-\frac{||\alpha_3||^2 + ||\alpha_1||^2}{2} + \tilde{\alpha}_{3_I} \alpha_1^I} 
    \sum_{s=0}^{\infty} (\tilde{\alpha}_{3K})^{s} \mathcal{S}^{-2, (J_1 ...J_{2l}),K_1 ...K_{s}, I_1 ...I_{s}} (\alpha_{1I})^{s} \,, \nonumber
\end{align}
which is more useful for matching. In the final expression the eigenvalue $g_{lm} = g_{lm} (\tilde{\alpha}_3,\alpha_1)$ of $\breve{g}_l$ was pulled out as the effective ``Wilson coefficient'' of graviton coupling to the spinning black hole, which can depend on the quantum number $m$ determined from the assignment of $J$ little group indices. For example, the $\breve{g}_l$ acting on $\hat{\mathcal{S}}$ defined by the rules
\begin{enumerate}
    \item Substitute all massive spinors with $J$ indices by the rule
\begin{align}
\begin{aligned}
    | \mathbf{3}^J]^{\dot\alpha} &\to | \mathbf{3}^J]^{\dot\alpha} + \zeta \left( \frac{p^\mu}{m^2} {S}_p^\nu \epsilon_{\mu\nu\lambda\delta} \left[ \frac{i}{2} \bar{\sigma}^{[\lambda} \sigma^{\delta]} \right]^{\dot\alpha}_{~\dot\beta} \right) | \mathbf{3}^J]^{\dot\beta}
    \\ | \mathbf{3}^J\rangle_{\alpha} &\to | \mathbf{3}^J\rangle_{\alpha} + \zeta \left( \frac{p^\mu}{m^2} {S}_p^\nu \epsilon_{\mu\nu\lambda\delta} \left[ \frac{i}{2} {\sigma}^{[\lambda} \bar{\sigma}^{\delta]} \right]_{\alpha}^{~\beta} \right) | \mathbf{3}^J\rangle_{\beta}
\end{aligned}
\end{align}
    where $S_p^\mu$ is the spin operator for the coherent spin states given in \eqref{spin operator}.
    \item Only keep the linear-in-$\zeta$ terms of the resulting expression.
\end{enumerate}
yield an effective Wilson coefficient $g_{lm} = \textsl{a} m$. We can expect from this example that $m^n$ dependence of $g_{lm}$ always appears as $(\textsl{a}m)^n$.

The $\mathcal{S}^{-2,J_{2l},K_{s},I_{s}}$ are spinor-bracket-valued coefficients encoding kinematic information, which we fix by comparing to the spheroidal harmonics. In the classical limit $\hbar \to 0$ the expression localises to the condition $\alpha_1 = \alpha_3$ due to the exponential coefficient in the front. We assume this condition when matching the ansatz \eqref{pre_tensor} to spheroidal tensors.

The expansion coefficients $\mathcal{S}^{-2,J_{2l},K_{s},I_{s}}$ reproduce the $a\omega$ expansion of the spheroidal harmonics
\begin{align}
\begin{gathered}
    \lim_{\hbar \to 0} (\tilde{\alpha}_{K})^{s} \mathcal{S}^{-2, (J_1 ...J_{2l}),K_1 ...K_{s}, I_1 ...I_{s}} (\alpha_{I})^{s} = (a \omega)^s {_{-2}}S^{(s)}_{lm}(\theta,\phi) \,,\\ 
    {_{-2}}S_{lm}(a\omega,\theta,\phi) = \sum_{s=0}^\infty (a \omega)^s {_{-2}}S^{(s)}_{lm}(\theta,\phi) \,,
\end{gathered}
\end{align}
and are constrained by the conditions
\begin{enumerate}
    \item The helicity weight for $\{ |2\rangle, |2] \}$ adds up to $h = -2$.
    \item The little group weight for $\{ |{\mathbf1}^I\rangle , |{\mathbf1}^I] \}$ adds up to $s$ (there are $s$ spinor brackets).
    \item The little group weight for $\{ |{\mathbf3}^I\rangle, |{\mathbf3}^I] \}$ adds up to $2l + s$.
\end{enumerate}
A general ansatz is given as
\begin{equation}\label{curl_S}
\begin{aligned}
    \mathcal{S}^{-2, (J_1 ...J_{2l}),K_1 ...K_{s}, I_1 ...I_{s}}
    =&\frac{1}{{_{-2}}N_{lm}}  \sum_{\substack{i+j+r=s \\ i,j,u,(r-u) \ge 0}}
    \bigg\{ c^{(r-u,u,i,j)}_{l,0} \langle 2{\mathbf3}^J \rangle^{l+2-r}  \lbrack 2{\mathbf3}^J \rbrack^{l-2} \langle 2{\mathbf3}^K \rangle ^{r} \langle {\mathbf3}^J {\mathbf1}^I \rangle  ^{r-u}\SB{{\mathbf3}^J {\mathbf1}^I }  ^{u} \\
    &+
    c^{(r-u,u,i,j)}_{l,1} \langle 2{\mathbf3}^J \rangle^{l+2}  \lbrack 2{\mathbf3}^J \rbrack^{l-2-r}  \SB{2{\mathbf3}^K } ^{r} \AB{{\mathbf3}^J {\mathbf1}^I }  ^{r-u}\SB{{\mathbf3}^J {\mathbf1}^I}  ^{u}  \bigg\}\\
    &
    \times ( \lbrack 2{\mathbf3}^{K} \rbrack \langle 2{\mathbf1}^{I} \rangle )^i (\langle 2{\mathbf3}^{K} \rangle  \lbrack 2{\mathbf1}^{I} \rbrack )^j(k_2 \cdot p_3)^{r}, 
\end{aligned}
\end{equation}
where symmetrisation over the little group indices are implicit, and the normalisation for the leading term ($s=0$) was fixed by matching to the Schwarzschild limit \eqref{3pt_amplitude}. The $c^{(r,i,j)}_{l,\#}$ coefficients are determined by matching to spheroidal harmonics in the classical limit of the ansatz, which do not yield a unique solution due to degenerate classical limit. 
We list the first two coefficients as examples below,
\begin{align*}
    \mathcal{S}^{-2, (J_1 ...J_{2l})} &= \frac{c^{(0,0,0,0)}_{l,0}}{{_{-2}}N_{lm}}\langle 2{\mathbf3}^{(J} \rangle^{l+2} \lbrack 2{\mathbf3}^{J)} \rbrack^{l-2} \,,
    \\ 
    \mathcal{S}^{-2, (J_1 ...J_{2l}),K,I} &= \frac{1}{{_{-2}}N_{lm}}
   \left(
  c^{(0,0,1,0)}_{l,0}\langle 2{\mathbf3}^{J} \rangle^{l+2} \lbrack 2{\mathbf3}^{J} \rbrack^{l-2} 
\SB{2{\mathbf3}^K}\AB{2{\mathbf1}^I} \right. \\
&\left.\quad\quad+
c^{(0,0,0,1)}_{l,0}\langle 2{\mathbf3}^{J} \rangle^{l+2} \lbrack 2{\mathbf3}^{J} \rbrack^{l-2} 
\AB{2{\mathbf3}^K}\SB{2{\mathbf1}^I}
\right)\\
&\quad\quad+\left(
c^{(1,0,0,0)}_{l,0}\langle 2{\mathbf3}^J \rangle^{l+1}  \lbrack 2{\mathbf3}^J \rbrack^{l-2} \langle 2{\mathbf3}^K \rangle \langle {\mathbf3}^J {\mathbf1}^I \rangle \right.\\
&\quad\quad+
c^{(0,1,0,0)}_{l,0}\langle 2{\mathbf3}^J \rangle^{l+1}  \lbrack 2{\mathbf3}^J \rbrack^{l-2} \langle 2{\mathbf3}^K \rangle \SB{{\mathbf3}^J {\mathbf1}^I}\\
&\left.\quad\quad+
c^{(1,0,0,0)}_{l,1}
\langle 2{\mathbf3}^J \rangle^{l+2}  \lbrack 2{\mathbf3}^J \rbrack^{l-3} \lbrack 2{\mathbf3}^K \rbrack \langle {\mathbf3}^J {\mathbf1}^I \rangle  
\right. \\
&\left.\quad\quad+
c^{(0,1,0,0)}_{l,1}
\langle 2{\mathbf3}^J \rangle^{l+2}  \lbrack 2{\mathbf3}^J \rbrack^{l-3} \SB{2{\mathbf3}^K} \SB{{\mathbf3}^J {\bf1}^I}  
\right)(k_2 \cdot p_3)\,,
\end{align*}
and the constraints on the coefficients $c^{(r-u,u,i,j)}_{l,\#}$ by matching to spheroidal harmonics are given in table \ref{l=2_coef} for $l=2$. 
\begin{table}
\renewcommand{\arraystretch}{1.5}
\begin{center}
    \begin{tabular}{ | c | c | } 
  \hline
  $(a\omega)^0$& $c^{(0,0,0,0)}_{2,0}=1$  \\ 
  \hline
  $(a\omega)^1$ & $c^{(0,0,0,1)}_{2,0}=\frac{-5}{9},\;c^{(0,0,1,0)}_{2,0}=\frac{-1}{9},\;c^{(0,1,0,0)}_{2,0}=\frac{-1}{9}+c^{(1,0,0,0)}_{2,0}\;$ \\ 
  \hline
  $(a\omega)^2$ & $c^{(0,0,0,2)}_{2,0}=\frac{2035}{15876},\;c^{(0,0,1,1)}_{2,0}=\frac{505}{3969},\;c^{(0,0,2,0)}_{2,0}=\frac{103}{15876},\;c^{(0,1,0,1)}_{2,0}=\frac{79}{441}+c^{(1,0,0,1)}_{2,0},$\\\quad&$\;c^{(0,1,1,0)}_{2,0}=\frac{85}{1323}+c^{(1,0,1,0)}_{2,0},\;c^{(0,2,0,0)}_{2,0}=\frac{76}{1323}-c^{(2,0,0,0)}_{2,0}+c^{(1,1,0,0)}_{2,0}$ \\ 
  \hline
\end{tabular}
\caption{Constraints on the non-zero coefficients $c^{(r-u,u,i,j)}_{l,\#}$ for $l=2$. 
}
\label{l=2_coef}
\end{center}
\end{table}
We can see that the coefficient $c^{(1,0,0,0)}_{2,0}$ is a free parameter not fixed by matching to spheroidal harmonics. 

In the case of opposite helicity $h=+2$, we define the corresponding kinematic factor $\mathcal{S}^{+2}_{J_{2l},K_s,I_s}$ as the complex conjugate 
\begin{equation}
\begin{gathered}
    \mathcal{S}^{+2}_{ (J_1 ...J_{2l}),K_1 ...K_{s}, I_1 ...I_{s}}
    =\left(\mathcal{S}^{-2, (J_1 ...J_{2l}),K_1 ...K_{s}, I_1 ...I_{s}}\right)^*,\\
    \lim_{\hbar \to 0} {(\alpha^K)}^s\mathcal{S}^{+2}_{ (J_1 ...J_{2l}),K_1 ...K_{s}, I_1 ...I_{s}}(\tilde{\alpha}^I)^s
    =\left({_{-2}}S^{(s)}_{lm}(\theta,\phi)\right)^*(a\omega)^s.
\end{gathered}
\end{equation}
In summary, we write the spin-weighted spheroidal tensors as
\begin{align}\label{Sphe_tensor}
\begin{aligned}
    \mathcal{A}^{\left<a\right>,-2,(J_{2l})} ( \alpha_1,\tilde{\alpha}_3) &=  m_1g_{lm} e^{-\frac{||\alpha_3||^2 + ||\alpha_1||^2}{2} + \tilde{\alpha}_{3_I} \alpha_1^I} 
    \sum_{s=0}^{\infty} (\tilde{\alpha}_{3K})^{s} \mathcal{S}^{-2, (J_{2l}), K_{s}, I_{s}} (\alpha_{1I})^{s} \,, \\
    \mathcal{A}^{\left<a\right>,+2}_{(J_{2l})}(\alpha_3,\tilde{\alpha}_1)
    &=
   m_1 \tilde{g}_{lm} e^{-\frac{||\alpha_3||^2 + ||\alpha_1||^2}{2} + \tilde{\alpha}_{1_I} \alpha_3^I}
     \sum_{s=0}^{\infty} 
    {(\alpha_3^K)}^s\mathcal{S}^{+2}_{ (J_{2l}),K_{s},I_{s}}(\tilde{\alpha}_1^I)^s \,,
\end{aligned}
\end{align}
where we have omitted repeated symmetrised indices for brevity. 
The effective ``Wilson coefficients'' $g_{lm}$ and $\tilde{g}_{lm}$ are not constrained to be equal, but parity relates the two amplitudes and imposes the condition $\Tilde{g}_{lm}=(g_{lm})^*$. 

\section{Matching to absorption cross section of spinning black hole}\label{sec:3}
From an on-shell point of view, an absorption process can be modelled as a massive particle with mass $m_1$ and spin $s$ absorbing a graviton and becoming a massive $m_3$ particle with spin $s'$, where $l \ll s$ is the orbital angular momentum quantum number of the graviton before being absorbed by particle $1$ and $s'$ lies in the range $s - l \le s' \le s + l$ due to angular momentum addition rules. Keeping track of all possible $s'$ can become cumbersome, but use of coherent spin states simplifies the calculations. We expect the difference of the two description to be negligible, similar to how we expect canonical ensembles to be equivalent to microcanonical ensembles in the thermodynamic limit. We have constructed the relevant expressions in the previous section as \eqref{Sphe_tensor}, with free parameters $g_{lm}$ and $\tilde{g}_{lm}$ to encode the information of the dynamics.

We attempt to fix these effective ``Wilson coefficients'' of the absorption three-point amplitudes \eqref{Sphe_tensor} by matching to black hole perturbation theory (BHPT) calculations of absorption. In BHPT, the absorption cross-section is defined as the ratio of the flux across the horizon to the incoming flux from infinity, which can be computed by solving the Teukolsky equation~\cite{Maldacena:1997ih,Ivanov:2022qqt}. While closed-form solutions are not known, the equations can be solved as a long wavelength expansion where $Gm_1\omega\ll 1$ (which implies $a\omega\ll 1$) is the expansion parameter. 
Matching the absorption cross-section determines the (integrated) absolute square of the effective Wilson coefficient $|g_{lm}|^2$, which can be used to determine how the Compton amplitude becomes corrected by absorption effects. We limit our attention to the case of $l = 2$, since this is the dominant contribution. The coefficients $|g_{2m}|^2$ determined by matching analytically continued BHPT results in the superextremal limit turns out to give contributions to the Compton amplitude that scales as $\mathcal{O}(G^1 a^5 \omega^5)$, in contrast to the non-spinning case where the contributions scale as $\mathcal{O}(G^6 m_1^5 \omega^5)$. This is because the length scale $Gm_1$ can be traded with the spin length $a$, which is also the reason why $R^2$ coupling may become relevant from $\mathcal{O}(S^4)$ for 2PM dynamics in the arbitrary-spin EFT~\cite{Bern:2022kto}.

\subsection{Calculation from on-shell spheroidal tensor}
We consider the inclusive process $p_1+(\omega,l,m,h)\rightarrow p_3$ where the massive particle of momentum $p_1$ absorbs a massless particle of energy $\omega$ and helicity $h$ with angular quantum numbers $\{ l, m \}$ and becomes a one-particle state of momentum $p_3$. We can write the probability as~\cite{Endlich:2016jgc,Aoude:2023fdm}
\begin{equation}
{_{h}}P_{lm}=\int_{0}^{\infty}dm_3^2\rho_l(m_3^2)\int \frac{d^3{p_3}}{(2\pi)^3}V\frac{\left|\ab{p_3}S \ak{\psi_\xi,\gamma_\zeta}\right|^2}{\AB{p_3p_3}\AB{\psi_\xi\psi_\xi}\AB{\gamma_\zeta\gamma_\zeta}},
    \label{eq:inc_prob}
\end{equation}
where $\rho_l(m_3^2)$ is the spectral function depending on $l$, $m_3$ is the mass of particle $3$, $V=\AB{p_3p_3}/2p_3^0$ is the phase space volume of $p_3$. 
To avoid subtleties with singular normalisation factors we use wavepackets to describe the incoming particles,
\begin{equation}\label{w_p}
    \begin{split}
\ak{\psi_{\xi}}=\int_{p_1}\psi_{\xi}(p_1)\ak{p_1} \,,\quad
\ak{\gamma_\zeta}=\int_{0}^{\infty}d\omega \gamma_{\zeta}(\omega)\ak{\omega,l,m,h} \,,
    \end{split}
\end{equation}
where the variables $\{ \xi, \zeta \}$ parametrise the wavepacket size and the states that are normalised by $\AB{\psi_\xi\psi_\xi}=\AB{\gamma_\zeta\gamma_\zeta}=1$. The Lorentz-invariant phase space measure is defined as
\begin{equation}
    \int_p:=\int \frac{dp^4}{(2\pi)^3}\delta(p^2-m_p^2)\theta(p^0) \,.
\end{equation}
The wavepacket states \eqref{w_p} behave as classical point-particles/monocromatic waves in the limit $\xi, \zeta \rightarrow
0$. Since the one-particle states used in amplitudes are plane wave states, we need to convert between plane wave one-particle states and spherical/spheroidal wave one-particle states. We generalise the conversion factor of ref.\cite{Aoude:2023fdm} to spheroidal waves as
\begin{equation}
    \AB{k_2,h|\omega,l,m,h'}=\frac{4\pi}{\sqrt{2\omega}}\delta^{h}_{h'}2\pi\delta(|\vec{k_2}|-\omega)\left({_{h'}}S_{lm}(a\omega,\hat{\bf{k}}_2)\right)^*,
\end{equation}
where we use spheroidal harmonics ${_{h}}S_{lm}$ in place of spherical harmonics ${_{h}}Y_{lm}$. We have implicitly set the time direction to be given by $p_1$ and the unit vector $\hat{\bf{k}}_2=\vec{k}_2/|\vec{k}_2|$ parametrises the angular directions. Inserting this conversion factor, the numerator of the integrand \eqref{eq:inc_prob} becomes
\begin{equation}
    \begin{split}
        \ab{p_3}S \ak{\psi_\xi,\gamma_\zeta}    =4\pi i\int_{p_1,k_2}\psi_\xi(p_1)\frac{\gamma_\zeta(k_2^0)}{\sqrt{2k_2^0}}(2\pi)^4\delta^{4}(p_1+k_2-p_{X_f})\mathcal{A}^{\left< a\right>}(p_3|p_1;k_2,h=-2)\left({_{-2}}S_{lm}(a\omega,\hat{\bf{k}}_2)\right)^*
    \end{split}
\end{equation}
for $h = -2$.
Integration of the wavefunction factors becomes trivial in the classical limit~\cite{Aoude:2023fdm}, and the inclusive probability \eqref{eq:inc_prob} becomes
\begin{equation}\label{onshell_cr_section}
    \begin{split}
       {_{-2}}P_{lm}
       &=C_0 \int d\Omega d\Omega'\; {_{-2}}S_{jm}(a\omega,\theta',\phi')\left({_{-2}}S_{jm}(a\omega,\theta,\phi)\right)^*\\
       &\quad\quad\times
       \int \frac{d^2 \tilde{\beta} d^2 \beta}{\pi^2} \left(\mathcal{A} ^{\left<a\right>,-2,({J_1...J_{2l}})}(\beta,\tilde{\alpha};\theta',\phi') \right)^* \mathcal{A}^{\left<a\right>,-2,({J_1...J_{2l}})} (\beta,\tilde{\alpha};\theta,\phi)
       ,
    \end{split}
\end{equation}
where we absorbed the integral factors into $C_0=\frac{\omega\rho_l(m_3^2)}{2m_1}$ with $m_3 \simeq m_1 + \hbar \omega$ because energy conservation condition in the classical limit localises $m_3$ to $m_1$.
To perform the angular integrals we first evaluate the second line,
\begin{equation} \label{abs_gluing}
    \begin{aligned}
   &\int \frac{d^2 \tilde{\beta} d^2 \beta}{\pi^2} \left(\mathcal{A} ^{\left<a\right>,-2,({J_1...J_{2l}})}(\beta,\tilde{\alpha};\theta',\phi') \right)^* \mathcal{A}^{\left<a\right>,-2,({J_1...J_{2l}})} (\beta,\tilde{\alpha};\theta,\phi)\\
   &=\int  \frac{d^2 \tilde{\beta} d^2 \beta}{\pi^2}  \mathcal{A} ^{\left<a\right>,+2}_{({J_1...J_{2l}})}(\alpha,\tilde{\beta};\theta',\phi') \mathcal{A}^{\left<a\right>,-2,({J_1...J_{2l}})} (\beta,\tilde{\alpha};\theta,\phi) \,,
\end{aligned}
\end{equation}
where we sum over the free indices $\{ J_i \}$. This is the advantage of the coherent spin formalism; the sum over all possible final state irreducible spin representations effectively reduces to the sum over a single spin representation. The leading contribution of \eqref{abs_gluing} in the $a\omega$ expansion for $l=2$ is
\begin{equation}\label{leading_aw}
  \begin{split}
        \sum_{m'}&\frac{m_1^2 |g_{2m'}|^2}{({_{-2}}N_{2m'})^2}\int  \frac{d^2 \tilde{\beta} d^2 \beta}{\pi^2}\left.
    e^{-\frac{||\alpha||^2 + ||\beta||^2}{2} + \tilde{\beta}_I \alpha^I}
    \SB{23_J}^{4} 
    \right|_{(\theta',\phi')}
     \left.e^{-\frac{||\alpha||^2 + ||\beta||^2}{2} + \tilde{\alpha}_I \beta^I}
    \AB{23^J}^{4}
    \right|_{(\theta,\phi)}\\
    &\quad\quad\quad=\sum_{m'} \frac{m_1^2 |g_{2m'}|^2e^{-||\alpha||^2}
    \SB{23_J}^{4} 
    \AB{23^J}^{4}
    }{({_{-2}}N_{2m'})^2}\int  \frac{d^2 \tilde{\beta} d^2 \beta}{\pi^2}
    e^{-||\beta||^2}
    e^{\tilde{\beta}_I \alpha^I+
     \tilde{\alpha}_I{\beta}^I
    }
  \end{split}
\end{equation}
where the square/angle spinor brackets are functions of $(\theta',\phi')$/$(\theta,\phi)$, their explicit forms being spin-weighted spherical harmonics
\begin{equation}
    \left( {_{-2}}Y_{2m'}(\theta',\phi')\right)^*=\frac{\SB{23_J}^{4} }{{_{-2}}N_{2m'}}\,,\;{_{-2}}Y_{2m'}(\theta,\phi)=\frac{\AB{23^J}^{4} }{{_{-2}}N_{2m'}} \,.
\end{equation}
The shift in $\beta$ variables trivialise the gaussian integral
\begin{equation}\label{co_var_shift}
    \Tilde{\beta}_I\rightarrow \Tilde{\beta}_I-\Tilde{\alpha}_{I},
    \;
    \beta_K\rightarrow\beta_K+\alpha_{K}
    ,
\end{equation}
and the integral \eqref{leading_aw} becomes
\begin{equation}
    \begin{split}
    &\sum_{m'}m_1^2 |g_{2m'}|^2
    \left({_{-2}}Y_{2m'}(\theta',\phi')\right)^*
    \left({_{-2}}Y_{2m'}(\theta,\phi)\right)
    \int  \frac{d^2 \tilde{\beta} d^2 \beta}{\pi^2}
    e^{-||\beta||^2}\\
    &=\sum_{m'}m_1^2 |g_{2m'}|^2
    \left({_{-2}}Y_{2m'}(\theta',\phi')\right)^*
    \left({_{-2}}Y_{2m'}(\theta,\phi)\right).
    \end{split}
\end{equation}
Therefore the leading $a\omega$ order inclusive probability for $l=2$, ${_{-2}}P^{(0)}_{2m}$, is
\begin{equation}\label{onshell_cr_section}
    \begin{split}
       {_{-2}}P^{(0)}_{2m}
       &=\frac{\omega\rho_2(m_3^2)}{2m_1} \int d\Omega d\Omega'\; \left({_{2}}Y_{jm}(\theta',\phi')\right)^*{_{2}}Y_{jm}(\theta,\phi)\\
       &\quad\quad\times\sum_{m'}
       m_1^2 |g_{2m'}|^2
       \left({_{-2}}Y_{2m'}(\theta',\phi')\right)^*
    \left({_{-2}}Y_{2m'}(\theta,\phi)\right)\\
    &=
    \frac{\rho_l(m_3^2)m_1\omega}{2}|g_{2m}|^2
    \end{split}
\end{equation}
The subsequent $a\omega$ expansion can be computed in a similar vein; we present the next-to-leading order calculations in appendix~\ref{Coherent_Gluing}.

\subsection{Matching to classical absorption cross section}
Using separation of variables, the spin $h$ Teukolsky scalar can be expanded in spin-weighted spheroidal harmonics
\begin{equation}
    {_h}\psi^P=e^{-i\omega t}\sum_{l,m}{_h}k^P_{lm}\;{_h}S_{lm}(a\omega,\theta,\phi){_h}R_{lm}(r) \label{eq:psi_sov}
\end{equation}
where parity $P=\pm1$ is the parity, ${_h}k^P_{lm}$ is a function of $a\omega$ and $Gm_1\omega$ associated with parity and the radial function ${_h}R_{lm}(r)$ is fixed by ingoing boundary conditions at the horizon. 
At infinity the radial function separates into the incoming part and the scattered part,
\begin{equation}\label{asym_rad}
    {_h}R_{lm}(r)\rightarrow {_h}B_{lm}^{\text{inc}}r^{-1}e^{-i\omega r_*}+{_h}B_{lm}^{\text{ref}}r^{-(2h+1)}e^{i\omega r_*}\;\text{as}\;r\rightarrow \infty
\end{equation}
where ${_h}B_{lm}^{\text{inc}}$ and ${_h}B_{lm}^{\text{ref}}$ are integration constants, and $r_*$ is the tortoise coordinate. To extract the scattering amplitude for a gravitational plane wave scattering off a black hole, one re-expresses the wave as a linear combination of metric perturbations, derived from the scalar functions ${_h} \psi^P$. As a result, one finds the (partial wave) phase shift ${_{h}}\delta^P_{lm}\in\mathbb{R}$ and transmission factor ${_{h}}\eta^P_{lm}\in\mathbb{R}$, given by  
\begin{equation}\label{phase_sh}
    {_{h}}\eta^P_{lm}e^{2i{_{h}}\delta^P_{lm}}=(-1)^{l+1}\frac{\text{Re}({_{h}C_{lm}(a\omega))+12iGm_1\omega P}}{(2\omega)^{2h}}\frac{{_{-h}}B_{lm}^{\text{ref}}}{{_{-h}}B_{lm}^{\text{inc}}}\,.
\end{equation}
where ${_{h}}C_{lm}(a\omega)$ is the Teukolsky-Starobinsky constant that only depends on $Gm_1\omega$ and the combination $(a\omega)$. The ratio of the integration constants is known to be computable using the method of matched asymptotics~\cite{Sasaki:2003xr}
\begin{equation}
    \omega^{2h}\frac{{_h}B_{lm}^{\text{ref}}}{{_h}B_{lm}^{\text{inc}}}= {\frac{1+ie^{i\pi\nu}\frac{K_{-\nu-1,h}}{K_{\nu,h}}}{1-ie^{-i\pi\nu}\frac{\sin{(\pi(\nu-h+i\epsilon))}}{\sin{(\pi(\nu+h-i\epsilon))}} \frac{K_{-\nu-1,h}}{K_{\nu,h}}} } {\frac{A^\nu_{-,h}}{A^\nu_{+,h}}e^{i\epsilon(2\ln(\epsilon)-(1-\kappa))}} \,,
\end{equation}
where $\epsilon=2 Gm_1\omega$ is the long wavelength expansion parameter and $\kappa=\sqrt{1-(a/Gm_1)^2}$ and the auxiliary parameter $\nu$ is a parameter introduced for the matching procedure.
The amplitudes are then given as~\cite{Futterman:1988ni,Glampedakis:2001cx}
\begin{align} \label{eq:f_amp_def}
    f(a\omega,\gamma,\theta,\phi)&=\frac{\pi}{i\omega}\sum_{l,m}{_{-h}}S_{lm}(a\omega,\gamma,\phi'){_{-h}}S_{lm}(a\omega,\theta,\phi)\sum_{P=\pm1}\left({_{h}}\eta^P_{lm}e^{2i{_{h}}\delta^P_{lm}}-1\right) \,,\\
    g(a\omega,\gamma,\theta,\phi)&=\frac{\pi}{i\omega}\sum_{l,m}{_{-h}}S_{lm}(a\omega,\gamma,\phi'){_{-h}}S_{lm}(a\omega,\pi-\theta,\phi)\sum_{P=\pm1}P(-1)^{l+m+h}\left({_{h}}\eta^P_{lm}e^{2i{_{h}}\delta^P_{lm}}-1\right) \,, \label{eq:g_amp_def}
\end{align}
The ${_{-h}}S_{lm}(a\omega,\gamma,\phi')$ factor parametrises the direction of the incoming plane wave, where the spin of the black hole $a$ is directed along the $z$-axis and the angular variables $\{ \gamma , \phi' \}$ determine the direction of the incoming momentum $\vec{k}_{in}$. The $-1$ factor in the sum is the remnant of subtracting the incoming plane wave contribution.\\
The elastic differential cross section is
\begin{equation}\label{diff_cross}
    \frac{d\sigma_{ela}}{d\Omega}=|f(a\omega,\gamma,\theta,\phi)|^2+|g(a\omega,\gamma,\theta,\phi)|^2.
\end{equation}
When $0\le{_{h}}\eta^P_{lm}<1$, there are complementary absorptive effects whose cross section is given by
\begin{equation}\label{absorption_cross}
    \begin{split}
        \sigma_{abs}&=\frac{4\pi^2}{\omega^2}\sum_{lm}\left|{_{-h}}S_{lm}(a\omega,\gamma,\phi')\right|^2 (1-{_{h}}\eta_{lm}^2) \,,
    \end{split}
\end{equation}
where we have drop the parity index and the absorption coefficient $_{h}\eta_{lm}$ is attributed to~\cite{Ivanov:2022qqt}

\begin{equation}
{_{h}}\eta^2_{lm}=\left|\frac{1+ie^{i\pi\nu}\frac{K_{-\nu-1,-h}}{K_{\nu,-h}}}{1-ie^{-i\pi\nu}\frac{\sin{(\pi(\nu+h+i\epsilon))}}{\sin{(\pi(\nu-h-i\epsilon))}} \frac{K_{-\nu-1,-h}}{K_{\nu,-h}}}\right|^2.
\end{equation}
From this expression, the absorptive partial wave cross section ${_{-h}} \sigma_{abs,lm}=(1-{_{h}}\eta_{lm}^2)$ in the long wavelength limit $Gm_1\omega\ll 1$ is given by ~\cite{Chia:2020yla,Ivanov:2022qqt} 
\begin{equation}
    \begin{split}
       {_{-h}} \sigma_{abs,lm}&=2(-1)^s\frac{(l-h)!(l+h)!}{(2l)!(2l+1)!}\left|2\omega(r_+-r_-)\right|^{2l+1} \text{Im}{{_{-h}}\mathcal{I}_{lm}},\\
     {_{-h}}\mathcal{I}_{lm}&=i(-1)^{h+1}P_+\frac{(l-h)!(l+h)!}{(2l)!(2l+1)!}\prod_{j=1}^{l}\left(|j+2iP_+|^2\right),\\
     P_+&=\frac{ma-2Gm_1r_+\omega}{r_+-r_-},
    \end{split}
\end{equation}
where $r_\pm=Gm_1\left(1\pm\sqrt{1-\left(a/Gm_1\right)^2}\right)$ is the position of the outer/inner horizon. Following the treatment of refs.\cite{Bautista:2021wfy,Bautista:2022wjf}, we analytically continue this solution to the superextremal limit $a/Gm_1\rightarrow\infty$ while keeping $a\omega\ll 1$,
leading to the substitutions
\begin{align}\label{sup_ex}
    \begin{split}
       \left|\omega( r_+-r_-)\right| &\rightarrow 2a\omega,\\ 
       P_+ &\rightarrow -\eta i\frac{m}{2}-Gm_1\omega,\\
        \text{Im}{_{-h}}\mathcal{I}_{lm} &\rightarrow (-1)^{h+1}\left(-Gm_1\omega\right)\frac{(l-h)!(l+h)!}{(2l)!(2l+1)!}\prod_{j=1}^{l}\left(j+\eta m\right)^2+O\left((Gm_1\omega)^2\right),\\
        {_{-h}}\sigma_{abs,lm} &\rightarrow 2Gm_1\omega\left(4a\omega\right)^{2l+1}\left(\frac{(l-h)!(l+h)!}{(2l)!(2l+1)!}\right)^2\prod_{j=1}^{l}\left(j+\eta m\right)^2+O\left((Gm_1\omega)^2\right),
    \end{split}
\end{align}
where $\eta=\pm1$ is the choice made for analytically continuing $\sqrt{1-(a/Gm_1)^2}\rightarrow \eta ia/Gm_1$.
Matching this partial wave cross section ${_{-2}}\sigma_{2m}$ to the inclusive probability ${_{-2}}P_{2m}$ computed in \eqref{onshell_cr_section}, we determine the effective Wilson coefficient up to spectral density
\begin{equation} \label{eq:eWC_sol}
    \rho_2(m_3^2)m_1^2\left|g_{2m}\right|^2
    =Gm^2_1(a\omega)^5\frac{64}{225}\left(1+\eta m\right)^2\left(2+\eta m\right)^2 \,.
\end{equation}
In addition, one can also analytically continue the elastic cross section \eqref{diff_cross} to superextremal limit $a/Gm_1\rightarrow\infty$. Its contribution to the same order of \eqref{sup_ex} is expected and not captured by our matching to the absorptive process.

\section{Compton amplitude of spinning black hole}\label{sec:4}
The gravitational Compton amplitude constructed from minimal coupling through na\"ive recursion is known to have spurious poles from quintic order in spin, leading to various proposals for its resolution~\cite{Chung:2018kqs,Chiodaroli:2021eug,Aoude:2022trd,Cangemi:2022bew}. For determining which of the proposals actually describe black holes, and to what extent the minimal coupling Compton amplitude describes spinning black holes, a series of studies based on BHPT was initiated~\cite{Bautista:2021wfy,Bautista:2022wjf} with a rather surprising conclusion; the minimal coupling Compton amplitude seems to capture black hole dynamics correctly up to quartic order in spin, but from the quintic order non-polynomial spin term of the form $|a| = \sqrt{- a_\mu a^\mu}$ should be present. The conclusion is surprising because non-polynomial operators are rarely considered in EFT calculations, since such operators signal a singularity around the perturbation point responsible for the non-analyticity.

The general view was that such a $|a|$ term is due to dissipative effects associated with the horizon,\footnote{At least during the workshop ``Amplifying Gravity at All Scales'' in July 2023.} since in BHPT calculations the term can be traced back to the term of the form $\sqrt{1-(a/Gm_1)^2}$, which becomes proportional to $\pm i |a|$ when analytically continued to superextremal limit where the matching to QFT amplitude is performed. The analytic continuation mixes real and imaginary parts of the solution, which are associated to conservative and dissipative effects respectively; therefore it was viewed that mixing between conservative and dissipative effects is responsible for the appearance of $|a|$-like terms in the amplitude. 

From the effective Wilson coefficient $|g_{2m}|^2$ determined in the last section, we compute the leading contributions to the spinning Compton amplitude from exchange of states associated to horizon absorption, for the opposite helicity or helicity-preserving configuration. Due to the explicit $m$ quantum number dependence of the effective Wilson coefficient $|g_{2m}|^2$, in gluing the three-point amplitude one needs to sum over $m$ along with integrating over the auxiliary variables for the coherent spin states. We then match onto the basis of Lorentz-invariant tensor structures for the gravitational Compton amplitude constructed in ref.\cite{Bautista:2022wjf}. The reconstructed Compton contribution scales as $\mathcal{O}(G^1 a^5 \omega^5)$ and indeed contains non-analytic terms of the form $|a|$, albeit the relative ratio of the coefficients between tensor structures turn out to be different from that given by ref.\cite{Bautista:2022wjf}.

For the same helicity or helicity-flipping configuration of the Compton amplitude we argue that the contribution consists of parity-even and parity-odd terms, which due to degeneracy cancels against each other and results in vanishing corrections. We remark that a similar argument for the non-spinning case was presented in ref.\cite{Jones:2023ugm}. This argument partly explains why the same helicity minimal coupling Compton amplitude was found to match the BHPT calculations~\cite{Bautista:2021wfy,Bautista:2022wjf}.

\subsection{Kinematic setup}

We use the following kinematic configuration to glue together two three-point amplitudes constructed in the previous section.
\begin{equation}\label{Compton_kine}
\begin{aligned}
    p_1^\mu &= 
    \begin{pmatrix}
	m_1\,, &0 \,, &0 \,, &0
    \end{pmatrix} \,,\\
    k_2^\mu &=  \omega
    \begin{pmatrix}
       1 \,, &\sin\gamma \,, &0 \,, &\cos\gamma
    \end{pmatrix}
    \,,\\
    k_3^\mu &= \frac{\omega m_1}{\omega+m_1-\omega \cos{(\gamma-\theta})}
    \begin{pmatrix}
	1 \,, &\sin{\theta} \,, &0 \,, &\cos{\theta}
    \end{pmatrix} \,,\\
    p_4^\mu &= p_1^\mu+k_2^\mu- k_3^\mu \,,\\
    p^\mu &= p_1^\mu +k_2^\mu \,.
\end{aligned}
\end{equation}
The particles 1 \& 4 correspond to spinning black holes and 2 \& 3 are gravitons, where 1 \& 2 are incoming and 3 \& 4 are outgoing. $p^\mu$ is the momentum carried by the intermediate exchanged state in the direct channel ($s$-channel). We assign helicity $+2$ to both gravitons, therefore the helicity assignment is $\{ h_2 = +2 , h_3 = - 2 \}$ in the all-incoming convention. The spin of the spinning black hole $a^\mu = (0,a \hat{z})$ is directed along the $z$-axis and the configuration is a restricted configuration where the three vectors $\{ \vec{a} , \vec{k}_2 , \vec{k}_3 \}$ lie on a plane, but the configuration contains enough information to perform functional reconstruction free of ambiguities. We have confirmed that inclusion of the azimuthal angle $\varphi$ dependence in $k_3$ is consistent with the Compton amplitude fixed without the $\varphi$ dependence.
\begin{figure}[htp]
\begin{center}
    \includegraphics[width=10cm]{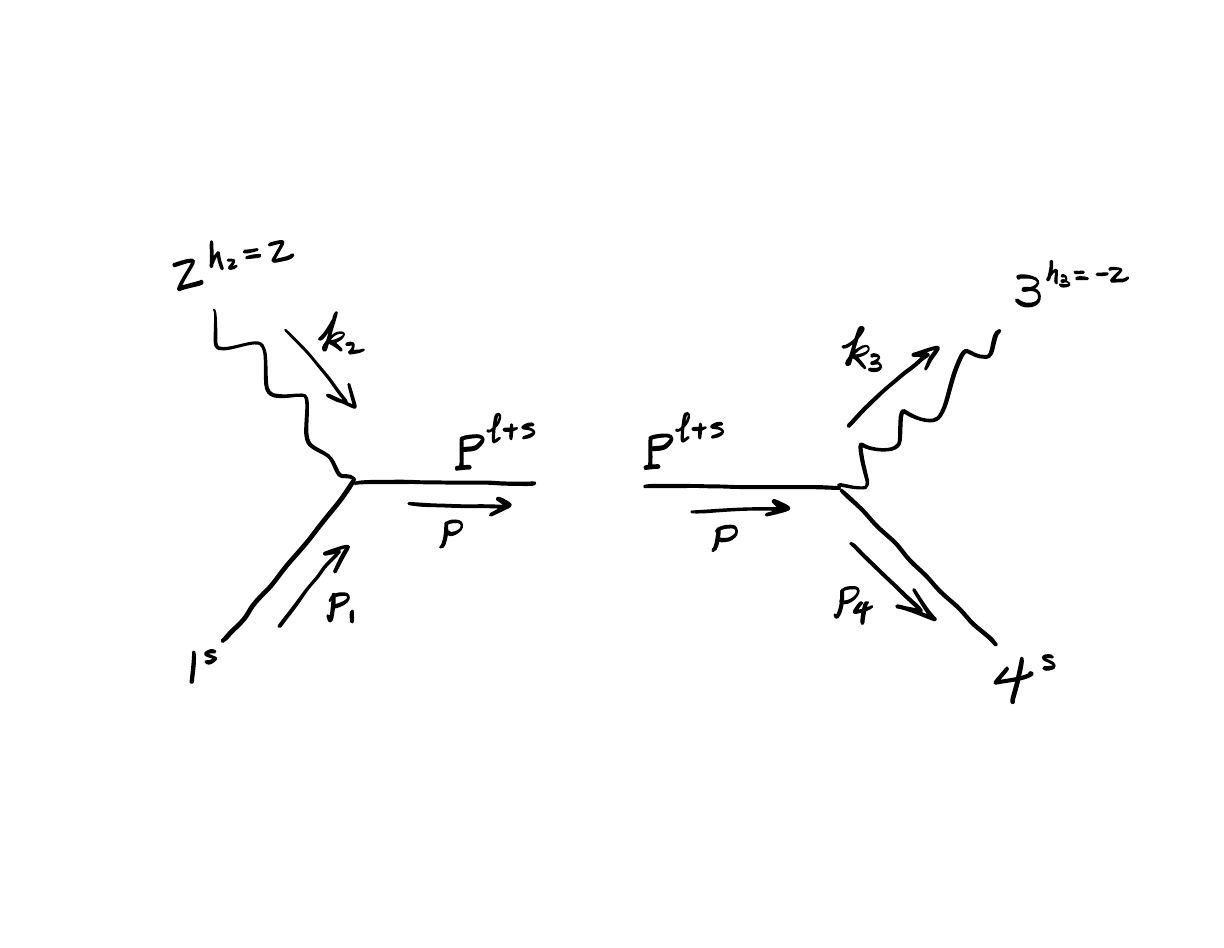}
    \caption{The Compton amplitude of ingoing massive particle $1^s$, massless particle $2^{h=2}$, intermediate particle $P^{l+s}$ and outgoing massless particle $3^h=-2$, massive particle $4^s$}
    \label{4pt}
    \end{center}
\end{figure}
\subsection{Compton amplitude from on-shell spheroidal tensor}
The contribution to the Compton amplitude from exchange of intermediate states associated to absorption can be computed by gluing two three-point amplitudes
\begin{align}
\begin{aligned} \label{compton}
    A^{abs}_{4,+-} &= \sum_l\int_{m_1^2}^{\infty}\frac{d(M^2)\rho_l(M^2)}{(p_1+k_2)^2-M^2}\int \frac{d^2 \tilde{\beta} d^2 \beta}{\pi^2}  \mathcal{A}_{3L} ^{\left<a\right>,+2,L_1...L_{2l}} (\alpha_1,\tilde{\beta})\mathcal{A}^{\left<a\right>,-2}_{3R,{L_1...L_{2l}}} (\beta,\tilde{\alpha}_4)
    \\ &\phantom{=asdf} + (k_2\leftrightarrow-k_3)
\end{aligned}
\end{align}
where $M$ is the rest mass of the exchanged state and $l$ is the orbital angular momentum of the absorbed graviton, the leading contribution coming from $l = 2$. The $(k_2 \leftrightarrow - k_3)$ on the second line denotes the crossed channel ($u$-channel) contribution.
Following ref.\cite{Jones:2023ugm}, we separate the $d (M^2)$ integral domain into ``heavy'' and ``light'' regions. The ``heavy'' modes correspond to the integration domain $M^2-m_1^2 \gg 2p_1\cdot k_2$, 
\begin{equation}\label{compton_heavy0}
    \begin{split}
         A^{abs,heavy}_{4,+-}&=\sum_l\int_{\mu}^{\infty}\frac{d(M^2)\rho_l(M^2)}{\Delta(M^2) 
         }\int \frac{d^2 \tilde{\beta} d^2 \beta}{\pi^2}  \mathcal{A}_{3L} ^{\left<a\right>,+2,L_1...L_{2l}} (\alpha_1,\tilde{\beta})\mathcal{A}^{\left<a\right>,-2}_{3R,{L_1...L_{2l}}} (\beta,\tilde{\alpha}_4) \\
         &\phantom{=asdfasdf} + (k_2\leftrightarrow-k_3) 
    \end{split}
\end{equation}
where $\mu$ is the light-heavy separation scale satisfying $\mu-m_1^2\gg 2p_1\cdot k_2$ and $\Delta(M^2)=m^2_1-M^2$. While ref.\cite{Jones:2023ugm} did not evaluate the heavy region integration and only treated the contribution as an effective contact operator, we choose to evaluate the integral since one of our goals is to understand whether the non-analytic $|a|$-type operators can have a quantum field theoretic origin.

We assume $G m_1 \omega \ll 1$ is sufficiently small so that there is an intermediate scale $G m_1 \omega'\ll 1$, $\omega \ll \omega'$, which contributes to the integral \eqref{compton_heavy0} at the lower end of the integration domain $M^2 \simeq m_1^2 + 2 m_1 \omega' \sim \mu$. We compute the contribution from this intermediate regime $G m_1 \omega' \ll 1$ based on the effective Wilson coefficient \eqref{eq:eWC_sol} computed in the long wavelength limit. Formally, \eqref{compton_heavy0} can be written as
\begin{equation}\label{compton_heavy}
    \begin{split}
    A^{abs,heavy}_{4,+-} &= \sum_{lm,s}Gm_1^2(a\omega)^{2l+1} \int_{\mu}^{\infty}dM^2\frac{\rho_l(M^2)m_1^2}{\Delta(M^2)}
   \left(
   F^{(s)}_{lm}(\gamma,\theta)(a\omega)^{s}
   +(k_2\leftrightarrow-k_3)\right) + R \,,
    \end{split}
\end{equation}
where $R$ is the remainder term not captured by the long wavelength approximation and $F^{(s)}_{lm}(\gamma,\theta)$ encodes angular dependence of the integrand at order $(a\omega)^{s}$ determined from the three-point amplitudes. The functions $F^{(s)}_{lm}(\gamma,\theta)$ can be computed order by order in $(a \omega)$ as presented in appendix \ref{Coherent_Gluing}. The leading term is the $l = 2$ contribution contributing at $(a\omega)^5$ order,
\begin{equation}\label{leading_heavy}
    \left. A^{abs,heavy}_{4,+-} \right|_{\text{Leading},l=2} = \int_{\mu}^{\infty} d (M^2) \frac{\rho_2(M^2)m_1^2}{\Delta(M^2)} \frac{|g_{2m}|^2}{({_{-2}}N_{2m})^2} \AB{3P^J}^4
        \SB{2P_J}^4 + (k_2\leftrightarrow-k_3) \,,
\end{equation}
where we have suppressed the remainder term at this order. The crossed channel contribution $( k_2 \leftrightarrow -k_3)$ becomes indistinguishable from the direct channel contribution in the classical limit. Using the explicit parametrisation of the kinematics \eqref{Compton_kine} and inserting the effective Wilson coefficient \eqref{eq:eWC_sol}, we find the heavy mode contribution as 
\begin{align}
    A^{abs,heavy}_{4,+-} &=  KG m_1^2 (a \omega)^5 \frac{5}{4} \left(16 (\eta -1)^2 (1-2 \eta )^2 \sin ^4\left(\frac{\gamma }{2}\right) \sin ^4\left(\frac{\theta }{2}\right)\right.\nonumber\\
    &+4 (\eta -2)^2 (\eta -1)^2 \sin ^2\left(\frac{\gamma }{2}\right) \sin (\gamma ) \sin ^2\left(\frac{\theta }{2}\right) \sin (\theta ) \nonumber
    \\ &+16 (\eta +1)^2 (2 \eta +1)^2 \cos ^4\left(\frac{\gamma }{2}\right) \cos ^4\left(\frac{\theta }{2}\right)+6 \sin ^2(\gamma ) \sin ^2(\theta )\nonumber\\
    &+\left.\frac{1}{4} (\eta +1)^2 (\eta +2)^2 \sin ^3(\gamma ) \csc ^2\left(\frac{\gamma }{2}\right) \sin ^3(\theta ) \csc ^2\left(\frac{\theta }{2}\right)\right) \,, \label{eq:Comp_heavy_res}
     \nonumber\\ K &:= \frac{32}{225\pi}\int_\mu^\infty \frac{d(M^2)}{\Delta(M^2)} \,,
\end{align}
where we only keep the leading order in $\hbar$ expansion.


Unlike the heavy mode contribution, the light mode contribution is subject to cancellation between the direct channel contribution and the crossed channel contribution in the classical limit. For example, the leading order contribution equivalent to \eqref{leading_heavy} vanishes for the light modes, 
\begin{equation}
     \left.A^{abs,light}_{4,+-}\right|_{\text{Leading},l=2}=\int_{m_1^2}^{\mu}dM^2\frac{\rho_2(M^2)m_1^2}{2p_1\cdot k_2}\frac{|g_{2m}|^2}{({_{-2}}N_{2m})^2}\AB{3P^J}^4
        \SB{2P_J}^4+(k_2\leftrightarrow-k_3)=0 \,,
\end{equation}
because the crossed channel contribution only differs by a sign from the direct channel in the classical limit due to $(p_1 \cdot k_2) = - (p_1 \cdot k_3) \times [1 + \mathcal{O}(\hbar)]$. So the contribution of light modes only appear at the next-to-leading order.

We match the $A^{abs,heavy}_{4,+-}$ of \eqref{leading_heavy} to the ansatz for spin tensor structures of the Compton used in ref.\cite{Bautista:2022wjf}, where we insert the effective ``Wilson coefficients'' \eqref{eq:eWC_sol} determined by matching to BHPT.
The ansatz is given as
\begin{equation}
    A^0_4\left(e^{(2w-k_3-k_2)\cdot a}+P_{\xi}\right)
\end{equation}
where $A_4^0=-\frac{\ASB{3|p_1|2}^4}{m_1^2(s-m_1^2)(u-m_1^2)t}$ is the non-spinning gravitational Compton amplitude, $w^\mu=\frac{p_1\cdot k_2}{p_1\cdot\epsilon_2}\epsilon_2^\mu$, $\epsilon_2^{\dot{\alpha}\alpha}=\frac{\sqrt{2}\sk{3}^{\dot{\alpha}}\ab{2}^{\alpha}}{\SB{32}}$ is the polarisation vector for positive helicity photon of momentum $k_2$, and $P_\xi$ denotes the contact terms as an expansion in $\xi=\frac{-1}{m_1^2t}(s-m_1^2)(u-m_1^2)$. The $P_{\xi}$ term is introduced to cancel the spurious pole from expanding the exponential, and we focus on the $\mathcal{O}(a^5)$ ansatz to match with \eqref{leading_heavy}, which is reproduced from ref.\cite{Bautista:2022wjf} below.\footnote{The convention for $k_3$ is different; we use outgoing $k_3$ in contrast to ingoing $k_3$ of ref.\cite{Bautista:2022wjf}.} 
\begin{gather} \label{J_Cont}
    \begin{aligned}
        & 32\pi Gm_1^2 A^0_4 \left(
     {\frac{1}{\xi}(w\cdot a)^4r^{(0)}_{|a|} + (w\cdot a)^2(w\cdot a-k_2\cdot a)(w\cdot a-k_3\cdot a)r^{(1)}_{|a|}} \right. \\
     &\left. {\phantom{\frac{1}{\xi} asdfasdfasdfasdfasdfasfasdf} + \xi(w\cdot a-k_2\cdot a)^2(w\cdot a-k_3\cdot a)^2r^{(2)}_{|a|}}
     \right) \,,
    \end{aligned}
    \\ r^{(j)}_{|a|}=c_2^{(j)}(k_2\cdot a+k_3\cdot a)+c_3^{(j)}(w\cdot a)+c_4^{(j)}|a|\omega \,.
\end{gather}
The $c_3^{(j)}$ coefficients must be tuned to cancel the spurious pole from expanding $e^{2w\cdot a}$, and this condition leads to the constraint $c_3^{(2)}=4/15-c_3^{(0)}+c_3^{(1)}$. The $c^{(j)}_4$ coefficients are responsible for the ill-understood $|a|$ terms of the Compton amplitude. Using the kinematics of \eqref{Compton_kine}, we find
\begin{equation}
   \begin{split}
        w\cdot a &=-a \omega \cos \left(\frac{\gamma +\theta }{2}\right) \sec \left(\frac{\gamma -\theta }{2}\right)\,,\\
        k_2\cdot a&=-a \omega\cos (\gamma )\,,\\
        k_3\cdot a&=-\frac{a \omega m_1 \cos (\theta )}{\omega+m_1-\omega \cos (\gamma -\theta )}\,,\\
        \xi&=-\csc ^2\left(\frac{\gamma -\theta }{2}\right)\,,\\
        A^{(0)}_4&=-\frac{1}{16} \sin ^4(\gamma -\theta ) \csc ^6\left(\frac{\gamma -\theta }{2}\right) \,,
        \end{split}  
    \end{equation}
with graviton energy $\omega$. 
We reorganise the ansatz \eqref{J_Cont} as
\begin{equation}\label{JT_contact}
32\pi G m_1^2 (a\omega)^5 \sum_{i=2}^{4}\sum_{j=0}^{2}c^{(j)}_{i}P_{i}^{(j)}(\theta,\gamma),
\end{equation}
where $P_{i}^{(j)}(\theta,\gamma)$ are angular functions associated to each coefficient $c_{i}^{(j)}$
. We use Fourier cosine series to match the ansatz $\eqref{JT_contact}$ to the absorption effect contribution \eqref{eq:Comp_heavy_res}
\begin{equation}\label{JT_F}
     P^{(j)}_i(\theta,\gamma)=\sum_{q,r}c^{(j)}_{i,qr}\cos{(q\theta+r\gamma)}
\end{equation}
where $q,r\in\mathbb{Z}$.
Matching \eqref{JT_F} to \eqref{eq:Comp_heavy_res} computed in the previous section, the coefficients of the ansatz \eqref{J_Cont} can be determined as
\begin{equation}\label{matching_JT}
    \begin{split}
        c_4^{(0)}&=\frac{45 K}{4 \pi },\;
        c_4^{(1)}=\frac{225 K}{16 \pi },\;
        c_4^{(2)}=\frac{195 K}{64 \pi },\\
        c_3^{(0)}&=\frac{4}{15}-\eta\frac{45 K}{4 \pi },\;
        c_3^{(1)}=-\eta\frac{315 K}{16 \pi },\;
        c_3^{(2)}=-\eta\frac{135 K}{16 \pi },\\
        c_2^{(j)}&=0,\;\;j=0,1,2.
    \end{split}
\end{equation}
We remark on the similarities and differences from the result presented in ref.\cite{Bautista:2022wjf}. The similarities are; a) the coefficients $c_2^{(j)}$ vanish, b) non-vanishing $c_4^{(i)}$ coefficients are consistent with existence of $|a|$ contributions. The difference is that the coefficients appear with different relative ratios, which may be due to remainder contributions $R$ neglected in \eqref{leading_heavy}. 

We can attempt a similar calculation for the helicity flipping ($++$ or $--$) configuration, which is limited by the fact that the absorption cross-section matching \eqref{eq:eWC_sol} only determines the \emph{absolute square} of the effective Wilson coefficient weighted by the spectral density $\rho_2(m_3^2) |g_{2m}|^2$, while we need the \emph{square} of the effective Wilson coefficient weighted by the spectral density $\rho_2(m_3^2) (g_{2m})^2$ for evaluating the analogue of \eqref{compton}. Interestingly, we can circumvent this problem using parity duality of BHPT; we can divide the sum in \eqref{compton} into parity-even (polar) interactions and parity-odd (axial) interactions, and show that the contributions cancel against each other. Note that an analogous conclusion was reached for the non-spinning case in ref.\cite{Jones:2023ugm}.

The transformation properties of the amplitude under parity determines the phase of the on-shell $S$-matrix element. A well-known example is the photon coupling to electric and magnetic charges; the electric coupling is parity-even and couples with a real coupling constant, while the magnetic coupling is parity-odd and couples with a purely imaginary coupling constant~\cite{Weinberg:1965rz}. This is the mechanism behind the on-shell electric-magnetic duality implemented by a phase rotation to the $x$-factor~\cite{Huang:2019cja, Emond:2020lwi, Kim:2020cvf}. This observation is relevant because BHPT predicts a duality between polar and axial perturbations of Kerr black holes~\cite{Li:2023ulk}; the absorption contribution \eqref{eq:eWC_sol} can be split into polar and axial contributions,
\begin{align}
    \rho_2(m_3^2) m_1^2 |g_{2m}|^2 = \rho_{2,p} (m_3^2) m_1^2 |g_{2m,p}|^2 + \rho_{2,a} (m_3^2) m_1^2 |g_{2m,a}|^2 \,,
\end{align}
where
\begin{align}
    \rho_{2,p} = \rho_{2,a} \,,\, |g_{2m,p}|^2 = |g_{2m,a}|^2 \,,
\end{align}
due to duality. Similar to the electric-magnetic duality, the effective Wilson coefficients are real/imaginary for the polar/axial modes, i.e.
\begin{align}
    (g_{2m,p})^2 = |g_{2m,p}|^2 \,,\, (g_{2m,a})^2 = - |g_{2m,a}|^2 \,.
\end{align}
These observations lead us to conclude that the numerator of the integrand \eqref{compton} for the same helicity configuration vanishes due to the duality,
\begin{align}
    \rho_2(m_3^2) m_1^2 (g_{2m})^2 = \rho_{2,p} (m_3^2) m_1^2 (g_{2m,p})^2 + \rho_{2,a} (m_3^2) m_1^2 (g_{2m,a})^2 = 0 \,,
\end{align}
therefore the Compton amplitude does not get modified by absorption effects for the same helicity configuration.
\section{Conclusion and outlook}\label{sec:5}
The coherent spin formalism provides a natural set-up for describing physics related to absorption/emission processes of spinning classical black holes, since the formalism naturally incorporates separation of scales of the spin degrees of freedom in the problem; the classical spin of the black hole scaling as $\mathcal{O}(\hbar^{-1})$ and the angular momentum carried by the massless quanta that scales as $\mathcal{O}(1)$. The usual irreducible representation approach to spin degrees of freedom inevitably mixes the scales and obscures the physics, and therefore is not well-suited for the problem of interest. The (spin-weighted) spheroidal harmonics\textemdash the special functions arising in BHPT relevant for describing absorption/emission\textemdash was shown to have a natural representation in terms of spinor variables in the coherent spin formalism, which would have been hard to find in the original spinor-helicity formalism since the spheroidal harmonics are \emph{not} irreducible representations of $SO(3)$. By matching to the absorption cross-section in the long wavelength approximation, we in turn computed the contribution of such absorptive process to the Compton amplitude. In the superextremal limit the corrections begin at 1 PM for $\mathcal{O}(S^5)$, and generate $|a|$ terms found in ref.\cite{Bautista:2022wjf}. An interesting observation was that the calculations provide an explanation for vanishing modifications to the helicity-flipping Compton amplitudes; from BHPT viewpoint there is a degeneracy in the polar and axial graviton perturbations, and while both channels contribute to the helicity-preserving configuration, their contribution cancel against each other in the helicity-flipping configuration.

The main difference from the non-spinning case~\cite{Kim:2020dif,Aoude:2023fdm} is that the effective contact-like interaction obatained after integrating out the exchanged excited states has the same $G$ counting as the leading contribution of the Compton amplitude, at least in the classical limit. This is because the length scale $Gm_1$ in the non-spinning case can be traded for the spin-length $|a|$ in the spinning case. The effective contact-like interactions from integrating out the intermediate states take the form of curvature squared ($R^2$) and higher derivative couplings, and to compensate for the dimensions of extra derivatives such contact-like interactions must be accompanied by factors of dimensions $[\text{Length}]^{4+n}$ where $n$ is the number of extra derivatives. For the non-spinning case such a factor can only be provided by the length scale of the horizon $Gm_1$, therefore such contact-like interactions cannot contribute to the $\mathcal{O}(G^1)$ terms of the Compton amplitude. For the spinning case the spin-length vector $a^\mu$ can be used instead of $Gm_1$, and the contact-like $R^2$ interactions can contribute to the $\mathcal{O}(G^1)$ Compton amplitude when accompanied by quartic or higher order spin factors~\cite{Bern:2022kto}. The same argument can be found in the context of post-Newtonian EFT for spin-tidal operators and their dimensionless Wilson coefficients~\cite{Kim:2021rfj,Kim:2022bwv,Levi:2022rrq}.

To conclude the paper, we comment on the effective contact-like curvature squared operators from an EFT point of view. Such contact-like operators arise from integrating out degrees of freedom we wish to ignore in the effective description, which in the usual EFT language is referred to as the high energy modes. Our computations show that the $|a|$-dependent terms of the $\mathcal{O}(G^1 S^5)$ Compton amplitude falls into this class of effective operators, where the $m$-dependence of the effective Wilson coefficient \eqref{eq:eWC_sol} seems to play a crucial role. On the other hand, their general presence is rather confusing from the viewpoint of BHPT; such contact-like operators have an interpretation as (adiabatic) tidal operators also known as (static) Love numbers, which for black holes are known to vanish~\cite{Damour:2009vw,Binnington:2009bb,Kol:2011vg,Chakrabarti:2013lua,Gurlebeck:2015xpa,Porto:2016zng,Hui:2020xxx,Charalambous:2021mea}. The vanishing of the static Love numbers has recently been attributed to hidden symmetries of the black hole spacetime~\cite{Charalambous:2021kcz, Hui:2021vcv}, which is likely to be an accidental symmetry of the static limit since dynamical Love numbers are known to be nonzero~\cite{Saketh:2023bul}. It would be interesting to study how these symmetries constrain the high energy modes to cancel out in the sum rule for tidal operators, although it is also possible that the cancellation is only realised at the level of thermal ensembles~\cite{Rothstein:2014sra}.


\section*{Acknowledgements}
YJC, TH and YTH,  are supported by MoST Grant No. 109-2112-M-002 -020 -MY3 and 112-2811-M-002 -054 -MY2. YJC would like to thank the hospitality of Max Planck Institute for Gravitational Physics (Albert Einstein Institute), Potsdam, during which part of the work was done.
The authors would like to thank Rafael Auode, Alexander Ochirov, Chia-Hsien Shen for insightful discussions.
JWK would like to thank Fabian Bautista, Lucile Cangemi, Maarten van de Meent, Paolo Picini, M.V.S. Saketh, Pratik Wagle for helpful discussions.
JWK would like to thank NORDITA for the hospitality during the workshop ``Amplifying Gravity at All Scales'', where part of this work was completed.

\appendix
\section{State sum for coherent spin states}\label{Coherent_Gluing}\label{app:a}
The resolution of the identity for coherent spin states is
\begin{align}
    \mathbb{I} &= \int \frac{d^2 \beta d^2 \tilde{\beta}}{\pi^2} | \beta \rangle \langle \beta | \,,
\end{align}
where we have suppressed the $SU(2)$ indices. When gluing the three-point amplitudes to construct the four-point, the ket/bra vectors split into the right/left subamplitudes $\mathcal{A}_{3L}^{L_1...L_{2l}}$/$\mathcal{A}_{3R}^{L_1...L_{2l}}$ and the $d^2 \beta d^2 \tilde{\beta}$ integral reduces to gaussian integrals,
\begin{align}\label{Gaussian_int}
    \int \frac{d \tilde{\beta} d \beta}{\pi^2}  \mathcal{A}_{3L} ^{L_1...L_{2l}} (\alpha_1,\tilde{\beta})\mathcal{A}_{3R,{L_1...L_{2l}}} (\beta,\tilde{\alpha}_4) \,,
\end{align}
where the gaussian factors are provided by the three-point amplitudes, 
\begin{align}
    \mathcal{A}_{3L}^{L_1...L_{2l}} (\alpha_1,\tilde{\beta}) &=
    e^{-||\alpha_1||^2/2}e^{-||\beta||^2/2}e^{\tilde{\beta}_I\mathbb{A}^I}\sum_{s}\tilde{\beta}_{I_1}\tilde{\beta}_{I_2}...\tilde{\beta}_{I_{s}} \mathcal{Q}^{I_1...I_{s} \,,{L_1...L_{2l}}}_{3L,s},\\
    \mathcal{A}_{3R,{L_1...L_{2l}}} (\beta,\tilde{\alpha_4}) &=
    e^{-||\alpha_4||^2/2}e^{-||\beta||^2/2}e^{\beta_I\mathbb{B}^I}\sum_{s}{\beta}_{I_1}{\beta}_{I_2}...{\beta}_{I_{s}}\mathcal{Q}_{3R,s}^{I_1,...,I_{s}}{}_{L_1...L_{2l}} \,,
\end{align}
where $\mathbb{A}^I,\mathbb{B}^I$ are functions of spinor variables and coherent state parameters $\{ \alpha_1, \tilde{\alpha_4} \}$, 
and the sums denote remainder terms where we keep the $\beta$ and $\tilde{\beta}$ dependence explicit. To do the gaussian integral we complete the squares through the redefinitions
\begin{align} \label{Coherent_Var_Shift}
\begin{aligned}
    \tilde{\beta_I}\rightarrow&  \tilde{\beta_I}-\mathbb{B}_I \,,\\
     {\beta_I}\rightarrow&  {\beta_I}+\mathbb{A}_I \,,
\end{aligned}
\end{align}
which reduces the integrals to the following family of master integrals
\begin{equation}
   Z_{I_1...I_{s},J_1...J_{s}}:= \frac{1}{\pi^2}e^{\frac{-1}{2}\left(||\alpha_1||^2+||\alpha_4||^2\right)-\mathbb{B}_I\mathbb{A}^I}\int d \tilde{\beta} d \beta \; e^{-|\beta|^2}\tilde{\beta}_{I_1}\tilde{\beta}_{I_2}...\tilde{\beta}_{I_{s}}{\beta}_{J_1}{\beta}_{J_2}...{\beta}_{J_{s}} \,.
\end{equation}
The master integrals can be evaluated using the generating function method, where we use the generating function
\begin{align}
    I_J=&\frac{1}{\pi^2}e^{\frac{-1}{2}\left(||\alpha_1||^2+||\alpha_4||^2\right)-\mathbb{B}_I\mathbb{A}^I}\int d \tilde{\beta} d \beta \; e^{-||\beta||^2+\tilde{\beta}_IJ^I+\beta_I\tilde{J}^I},\\
    =&e^{\frac{-1}{2}\left(||\alpha_1||^2+||\alpha_4||^2\right)-\mathbb{B}_I\mathbb{A}^I}e^{\tilde{J}^I\epsilon_{IK}J^K},
\end{align}
to get the master integrals
\begin{equation}\label{Inte_Result}
\begin{aligned}
    Z_{I_1...I_{s},K_1...K_{s}}=&\frac{\delta}{\delta J^{I}}...\frac{\delta}{\delta {J}^{I_{s}}}\frac{\delta}{\delta \tilde{J}^{K_1}}...\frac{\delta}{\delta \tilde{J}^{K_{s}}}I_J\left.\right|_{J=\tilde{J}=0}\\
    =&
    e^{\frac{-1}{2}\left(||\alpha_1||^2+||\alpha_4||^2\right)-\mathbb{B}_I\mathbb{A}^I}\sum_{\sigma}\left(
    \epsilon_{I_1\sigma\left(K_1\right)}
    \epsilon_{I_2\sigma\left(K_2\right)}\times...\times \epsilon_{I_{s}\sigma\left(K_{s}\right)}
    \right) \,,
\end{aligned}
\end{equation}
where $\epsilon_{IK}$ is the Levi-civita tensor and $\sigma$ stands for all possible permutations. Based on the evaluation procedure, we can separate the results of the integration \eqref{Gaussian_int} into three parts as
\begin{enumerate}
    \item Contributions from the variable shift \eqref{Coherent_Var_Shift}
     \begin{equation}
         \begin{split}
         & \mathcal{A}_{3L}^{L_1...L_{2l}} (\alpha_1,-\mathbb{B}) \mathcal{A}_{3R,{L_1...L_{2l}}} (\mathbb{A},\tilde{\alpha}_4)\\
         &=e^{\frac{-1}{2}\left(||\alpha_1||^2+||\alpha_4||^2\right)-\mathbb{B}_I\mathbb{A}^I}\sum_{s_1,s_2}
         (-1)^{s_1} \mathbb{B}_{I_1}...\mathbb{B}_{I_{s_1}}\mathcal{Q}^{I_1,...,I_{s_1},{L_1...L_{2l}}}_{3L,s_1}   \mathbb{A}_{K_1}...\mathbb{A}_{K_{s_2}}\mathcal{Q}^{K_1,...,K_{s_2}}_{3R,s_2,{L_1...L_{2l}}},
     \end{split}
     \end{equation}
    \item 
    Contributions from gaussian integration of $\tilde{\beta}, \beta$ polynomials \eqref{Inte_Result}
    \begin{align}
        e^{\frac{-1}{2}\left(||\alpha_1||^2+||\alpha_4||^2\right)-\mathbb{B}_I\mathbb{A}^I}\sum_{s}\mathcal{Q}^{I_1,...,I_{s},{L_1...L_{2l}}}_{3L,s}\epsilon_{I_1K_1}...\epsilon_{I_{s}K_{s}}\mathcal{Q}^{(K_1,...,K_{s})}_{3R,s,{L_1...L_{2l}}} \,,
    \end{align}
    \item Contributions from mixing of two effects, e.g.
    \begin{align}
        e^{\frac{-1}{2}\left(||\alpha_1||^2+||\alpha_4||^2\right)-\mathbb{B}_I\mathbb{A}^I}\sum_{s}\mathcal{Q}^{I_1,...,I_{s},{L_1...L_{2l}}}_{3L,s}(-\mathbb{B}_{I_1}\mathbb{A}_{K_1})...\epsilon_{I_{s}K_{s}}\mathcal{Q}^{(K_1,...,K_{s})}_{3R,s,{L_1...L_{2l}}} \,,
    \end{align}
\end{enumerate}
Of the three contributions, only the first part survives under classical limit ($a\rightarrow\infty$ with finite $a \omega$); the exponents scale as $\mathbb{A},\mathbb{B}\propto \sqrt{a}$ and the remainder terms scale as $\mathcal{Q}^{I_1,...,I_{s_1},{L_1...L_{2l}}}_{3,s}\propto \omega^{s/2} $, thus the integral \eqref{Gaussian_int} becomes
\begin{equation}\label{cl_glu}
    \begin{split}
        &\int \frac{d \tilde{\beta} d \beta}{\pi^2}  \mathcal{A}_{3L} ^{L_1...L_{2l}} (\alpha_1,\tilde{\beta})\mathcal{A}_{3R,{L_1...L_{2l}}} (\beta,\tilde{\alpha}_4)\\
        &\phantom{a} =e^{\frac{-1}{2}\left(||\alpha_1||^2+||\alpha_4||^2\right)-\mathbb{B}_I\mathbb{A}^I}\mathcal{Q}_{3L}^{L_1...L_{2l}} (\alpha_1,-\mathbb{B}) \mathcal{Q}_{3R,{L_1...L_{2l}}} (\mathbb{A},\tilde{\alpha}_4) \times \left[ 1 + O(a^{n_a}\omega^{n_b}) \right]
    \end{split}
\end{equation}
with $n_b>n_a\geq 0$. In the case of seed amplitudes \eqref{Sphe_tensor} for the four-point computation considered in the main text, the arguments of the the exponents $\mathbb{A},\mathbb{B}$ are $\alpha,-\tilde{\alpha}$ and \eqref{cl_glu} can be computed by simply replacing $\beta,\Tilde{\beta}$ with $\alpha,\tilde{\alpha}$,
\begin{equation}
    e^{\frac{-1}{2}\left(||\alpha_1||^2+||\alpha_4||^2\right)+\tilde{\alpha}_{4I}\alpha_1^I}\mathcal{A}_{3L}^{L_1...L_{2l}} (\alpha_1,\tilde{\alpha}_4) \mathcal{A}_{3R,{L_1...L_{2l}}} (\alpha_1,\tilde{\alpha}_4).
\end{equation}

As an example, we evaluate the case $l=2$ up to first $(a\omega)$ correction order. The seed left and right amplitudes are\footnote{We let $c^{(0,0,0,0)}_{2,0}=1$ for simplicity and denote $c^{(1,0,0,0)}_{2,0}=c_1,c^{(0,1,0,0)}_{2,0}=c_2,c^{(0,0,1,0)}_{2,0}=c_3,c^{(0,0,0,1)}_{2,0}=c_4$.}
\begin{align}
    \begin{split}
    \mathcal{A}_{3L}^{\left<a\right>,+2,(J_1J_2J_3J_4) }&=\frac{m_1g^*_{2m}}{{_{-2}}N_{2m}}e^{\frac{-||\beta||^2}{2}-\frac{||\alpha_1||^2}{2}}e^{\Tilde{\beta}_I\alpha_{1}^{I}}
    \\
    &\left(
    \SB{2{\bf p}^J}^4 
    +c_1\SB{2{\bf p}^J}^3\Tilde{\beta}_K\SB{2{\bf p}^K} \SB{{\bf p}^J{\bf1}^I} \alpha_{1I}(k_2\cdot p)\right.\\
    &+c_2\SB{2{\bf p}^J}^3\Tilde{\beta}_K\SB{2{\bf p}^K}\left(\AB{{\bf p}^J{\bf1}^I}\right)\alpha_{1I}(k_2\cdot p)\\
    &+c_3\SB{2{\bf p}^J}^4 \Tilde{\beta}_K\AB{2{\bf p}^K}\SB{2{\bf1}^I}\alpha_{1I} + \left. c_4\SB{2{\bf p}^J}^4\Tilde{\beta}_K\SB{2{\bf p}^K}\AB{2{\bf1}^I}\alpha_{1I}\right) \,,
    \end{split} \\
    \begin{split}
    \mathcal{A}_{3R}^{\left<a\right>,-2,(J_1J_2J_3J_4) }&=\frac{m_1g_{2m}}{{_{-2}}N_{2m}}e^{\frac{-||\beta||^2}{2}-\frac{||\alpha_4||^2}{2}}e^{\Tilde{\alpha}_{4K}\beta^{I}}
    \\
    &\left(
    \AB{3{\bf p}^J}^4 
    +c_1\AB{3{\bf p}^J}^3{\beta}_K\AB{3{\bf p}^K}\AB{{\bf p}^J{\bf4}^I}\Tilde{\alpha}_{4I}(k_3\cdot p)\right.\\
    &+c_2\AB{3{\bf p}^J}^3{\beta}_K\AB{3{\bf p}^K}\SB{{\bf p}^J{\bf4}^I}\Tilde{\alpha}_{4I}(k_3\cdot p)\\
    &+c_3\AB{3{\bf p}^J}^4{\beta}_K\SB{3{\bf p}^K}\AB{3{\bf4}^I}\Tilde{\alpha}_{4I}+\left.
 c_4\AB{3{\bf p}^J}^4{\beta}_K\AB{3{\bf p}^K}\SB{3{\bf4}^I}\Tilde{\alpha}_{4I}
    \right) \,.
\end{split}
\end{align}
Gluing the seed amplitudes, we get
\begin{equation}\label{gluing}
    \begin{aligned}
        &\int d\beta d\Tilde{\beta}\mathcal{A}_{3L}^{\left<a\right>,+2}{}_{(J_1J_2J_3J_4)}\mathcal{A}_{3R}^{\left<a\right>,-2,(J_1J_2J_3J_4)}\\
        &=
        \frac{m^2_1|g_{2m}|^2}{({_{-2}}N_{2m})^2}
        e^{\frac{-1}{2}\left(||\alpha_1||^2+||\alpha_4||^2\right)+\tilde{\alpha}_{4I}\alpha_1^I}\left(\AB{3{\bf p}^J}^4
        \SB{2{\bf p}_J}^4\right.+\\
        &
        \quad\quad\left.
        M_{1L,(J_1J_2J_3J_4)}\AB{3{\bf p}^J}^4+
        \SB{2{\bf p}_J}^4 M_{1R}^{(J_1J_2J_3J_4)}
        +O((a\omega)^2)\right)
    \end{aligned}
\end{equation}
where
\begin{equation}\label{M_nx}
    \begin{split}
     M_{1L}^{(J_1J_2J_3J_4)}=&c_1\SB{2{\bf p}^J}^3\Tilde{\alpha}_{4K}\SB{2{\bf p}^K}\left(\SB{{\bf p}^J{\bf1}^I}\right)\alpha_{1I}(k_2\cdot p)\\
    &+c_2\SB{2{\bf p}^J}^3\Tilde{\alpha}_{4K}\SB{2{\bf p}^K}\left(\AB{{\bf p}^J{\bf1}^I}\right)\alpha_{1I}(k_2\cdot p)\\
    &+c_3\SB{2{\bf p}^J}^4\left(\Tilde{\alpha}_{4K}\AB{2{\bf p}^K}\SB{2{\bf1}^I}\alpha_{1I}\right)+c_4\SB{2{\bf p}^J}^4\left(\Tilde{\alpha}_{4K}\SB{2{\bf p}^K}\AB{2{\bf1}^I}\alpha_{1I}\right) \,,\\
    M_{1R}^{(J_1J_2J_3J_4)}=
    &c_1\AB{3{\bf p}^J}^3{\alpha}_{1K}\AB{3{\bf p}^K}\AB{{\bf p}^J{\bf4}^I}\Tilde{\alpha}_{4I}(k_3\cdot p)\\
    &+c_2\AB{3{\bf p}^J}^3{\alpha}_{1K}\AB{3{\bf p}^K}\SB{{\bf p}^J{\bf4}^I}\Tilde{\alpha}_{4I}(k_3\cdot p)\\
    &+c_3\AB{3{\bf p}^J}^4{\alpha}_{1K}\SB{3{\bf p}^K}\AB{3{\bf4}^I}\Tilde{\alpha}_{4I}+
 c_4\AB{3{\bf p}^J}^4{\alpha}_{1K}\AB{3{\bf p}^K}\SB{3{\bf4}^I}\Tilde{\alpha}_{4I} \,.
    \end{split}
\end{equation} 
In the final step we use the kinematic setup \eqref{Compton_kine} and set $\alpha_1=\alpha_4$ to compute the integrand of \eqref{compton_heavy}, which to first subleading order becomes
\begin{equation}
    \begin{split}
        Gm^2_1(a\omega)^5F^{(0)}_{2,m}(\gamma,\theta)&=\frac{|g_{2m}|^2}{({_{-2}}N_{2m})^2}
        \AB{3{\bf p}^J}^4
        \SB{2{\bf p}_J}^4 \,,\\
        Gm^2_1(a\omega)^6 F^{(1)}_{2,m}(\gamma,\theta)&=\frac{|g_{2m}|^2}{({_{-2}}N_{2m})^2}\left(
        M_{1L,(J_1J_2J_3J_4)}\AB{3{\bf p}^J}^4+
        \SB{2{\bf p}_J}^4 M_{1R}^{(J_1J_2J_3J_4)}\right) \,.
    \end{split}
\end{equation}
\section{Equal-mass Compton amplitude from BCFW}\label{equalm_Com}
We use the following BCFW shifts where $\{ p_1,k_2 \}$ are incoming and $\{ k_3,p_4 \}$ are outgoing,
 \begin{equation}
    \begin{split}
        \hat{\lambda}_2=\lambda_2+z \lambda_3,\;\hat{\tilde{\lambda}}_3=\tilde{\lambda}_3+z\tilde{ \lambda}_2
    \end{split}
\end{equation}
 \begin{equation}
    \begin{split}
        \hat{k}_2=k_2+zq ,\;\hat{{k}}_3=k_3+zq
    \end{split}
\end{equation}
where $q=\lambda_3\tilde{ \lambda}_2$.\footnote{The other choice of spinor shifts do not yield sensible results.} We use the following minimal coupling amplitudes~\cite{Arkani-Hamed:2017jhn} as the seed amplitudes
\begin{equation}
M_{3L,min}^{s,+,\{I,J\}}({\bf1}^s,2^{+2},{\bf p}^s)=g\frac{x^2_{12}}{m^{2s}}\AB{{\bf p}^J{\bf 1}^I}^{2s},    
\end{equation}
\begin{equation}
M_{3R,min}^{s,-,\{I,J\}}({\bf p}^s,3^{-2},{\bf4}^s)=g(-1)^{2s}\frac{x^{-2}_{34}}{m^{2s}}\SB{{\bf 4}^J{\bf p}^I}^{2s}    
\end{equation}
where $g$ is the coupling constant and
\begin{equation}\label{x_factor}
    x_{12}=\frac{\SAB{2|p_1|3}}{m\AB{23}},\;x^{-1}_{34}=\frac{\ASB{3|p_1|2}}{m\SB{32}} \,,
\end{equation} 
are the $x$-factors. The BCFW recursion results in the Compton amplitude
\begin{equation}\label{s_Com}
    M_4({\bf1}^s,2^{+2},3^{-2},{\bf4}^s)= { -g^2}\frac{\ASB{{3}|p_1|2}^4}{m^2(s-m^2)(u-m^2)t}\left(\frac{\AB{{\bf4}^J3}\SB{{\bf1}^I2}+\AB{{\bf1}^I3}\SB{{\bf4}^J2}}{\ASB{3|p_1|2}}\right)^{2s} \,,
\end{equation}
where the amplitude develops spurious pole for $s>2$. 
The same computation can be applied to the minimal coupling coherent-spin amplitudes~\cite{Aoude:2021oqj},
\begin{equation}\label{ecoh_3pt}
    \begin{split}
        M^+_{3L}(\alpha_1,\tilde{\beta})&=gx^2_{12}e^{-(||\alpha_1||^2+||\beta||^2)/2}e^{\left(\frac{1}{m}\tilde{\beta}_I\AB{{\bf p}^I{\bf1}_J}\alpha_{1}^J\right)} \,,\\
        M^-_{3R}(\beta,\tilde{\alpha_4})&=gx^{-2}_{34}e^{-(||\alpha_4||^2+||\beta||^2)/2}e^{\left(\frac{1}{m}{\beta}_I\SB{{\bf p}^I{\bf4}_J}\tilde{\alpha}_{4}^J\right)} \,,
    \end{split}
\end{equation}
where we sum over intermediate states using \eqref{cl_glu}. The resulting Compton amplitude is
\begin{equation}\label{eqcoh_com}
    \begin{split}
    M_4(\alpha_1,\Tilde{\alpha}_4)&=-\sum_a\text{Res}_{z_a}\left(\frac{\int d\beta d\Tilde{\beta}\hat{M}^+_{3L}(\alpha_1,\tilde{\beta})\hat{M}^-_{3R}(\beta,\tilde{\alpha_4})}{z}\right)\\
    &={ -g^2}\frac{\ASB{{3}|p_1|2}^4}{m^2(s-m^2)(u-m^2)t}e^{\frac{-1}{2}\left(||\alpha_1||^2+||\alpha_4||^2\right)}e^{\frac{-{\alpha_1^I}\ASB{{\bf1}_{I}|\hat{p}|{\bf4}^J}\Tilde{\alpha}_{4J}}{m^2}}\\
    &={ -g^2}\frac{\ASB{{3}|p_1|2}^4}{m^2(s-m^2)(u-m^2)t}e^{\frac{-1}{2}\left(||\alpha_1||^2+||\alpha_4||^2\right)}e^{{\alpha_{1}^I}\left(\frac{\AB{{\bf4}^J3}\SB{{\bf1}_I2}+\AB{{\bf1}_I3}\SB{{\bf4}^J2}}{\ASB{3|p_1|2}}\right)\Tilde{\alpha}_{4J}}
    \end{split}
\end{equation}
which is reminiscent of the exponentiated form of the Compton amplitude~\cite{Guevara:2018wpp,Bautista:2019tdr,Aoude:2020onz,Chen:2021kxt}. Apart from obtaining \eqref{eqcoh_com} from coherent-spin amplitude \eqref{ecoh_3pt}, one can arrive at the same result by resumming the spin-$s$ Compton amplitude \eqref{s_Com} using coherent spin variables $\alpha_1,\Tilde{\alpha}_4$. 

\bibliography{refs}
\bibliographystyle{JHEP}

\end{document}